%% file: main.tex
\documentclass[largeformat]{interact}

\usepackage{epstopdf}% To incorporate .eps illustrations using PDFLaTeX, etc.
\usepackage[caption=false]{subfig}% Support for small, `sub' figures and tables
\usepackage{float}
\usepackage{xcolor}
\usepackage{caption}
\usepackage{microtype}
\usepackage{array}
\usepackage{wrapfig}
\usepackage{hyperref}
\usepackage{amsfonts}
\usepackage{booktabs}
\usepackage{enumitem}

\usepackage[natbibapa,nodoi]{apacite}% Citation support using apacite.sty. Commands using natbib.sty MUST be deactivated first!
\renewcommand\bibliographytypesize{\fontsize{10}{12}\selectfont}% To set the list of references in 10 point font using apacite.sty. Commands using natbib.sty MUST be deactivated first!

\theoremstyle{plain}% Theorem-like structures provided by amsthm.sty

\theoremstyle{definition}

\theoremstyle{remark}

\DeclareUnicodeCharacter{200E}{}

\def\markup{0} % change between 0 and 1 to switch between clean and marked-up versions
\if\markup 1
\usepackage{soul}
\newcommand{\rv}[1]{{\leavevmode\color{black}#1}}
\newcommand{\rrv}[1]{{\leavevmode\color{blue}#1}}

\else
\newcommand{\rv}[1]{#1}
\newcommand{\rrv}[1]{#1}
\newcommand{\st}[1]{}
\fi

\begin{document}

\articletype{Original Research}% Specify the article type or omit as appropriate

% \title{Taylor \& Francis \LaTeX\ template for authors (\textsf{Interact} layout + American Psychological Association reference style)}

% \author{
% \name{A.~N. Author\textsuperscript{a}\thanks{CONTACT A.~N. Author. Email: latex.helpdesk@tandf.co.uk} and John Smith\textsuperscript{b}}
% \affil{\textsuperscript{a}Taylor \& Francis, 4 Park Square, Milton Park, Abingdon, UK; \textsuperscript{b}Institut f\"{u}r Informatik, Albert-Ludwigs-Universit\"{a}t, Freiburg, Germany}
% }

\title{Understanding How Older Adults Comprehend COVID-19 Interactive Visualizations via Think-Aloud Protocol}

\author{\name{Mingming Fan\textsuperscript{a,b}\thanks{M. Fan is the corresponding author. Email: mingmingfan@ust.hk. Y. Wang and Y. Xie contributed equally.} and 
Yiwen Wang\textsuperscript{c} and 
Yuni Xie \textsuperscript{c} and
Franklin Mingzhe Li \textsuperscript{d} and
Chunyang Chen \textsuperscript{e}}
\affil{\textsuperscript{a}The Hong Kong University of Science and Technology (Guangzhou), China;\textsuperscript{b}The Hong Kong University of Science and Technology, Hong Kong SAR China;\textsuperscript{c}Rochester Institute of Technology, USA;\textsuperscript{d}Carnegie Mellon University, USA;\textsuperscript{e}Monash University, Australia}}

\maketitle

\begin{abstract}
% Visualizations have been widely used in live dashboards and websites to show dynamically evolving COVID-19 information (e.g., new and accumulative cases). Although visualizations can help older adults deal with age-related declines, older adults often encounter difficulties when reading visualizations. Little is known about how older adults would read COVID-19 visualizations and what challenges they might encounter. Furthermore, as COVID-19 visualizations tend to be interactive, it is unknown what challenges older adults might face with different interaction techniques adopted by such interactive visualizations. Motivated by these questions, we adopted a user-centered approach by inviting older adults to evaluate COVID-19 interactive visualizations. We conducted an online think-aloud usability testing with 18 older adults who interacted with COVID-19 interactive visualizations to complete tasks while verbalizing their thought processes. We further interviewed them to understand their experiences. From think-aloud data, we identified five thought processes representing how older adults comprehended the visualizations and uncovered the challenges they encountered with these thought processes. Furthermore, we also identified the challenges they encountered with seven common types of interaction techniques adopted by the visualizations. Based on the findings, we present design guidelines for making interactive visualizations more accessible to older adults.
Older adults have been hit disproportionally hard by the COVID-19 pandemic. One critical way for older adults to minimize the negative impact of COVID-19 and future pandemics is to stay informed about its latest information, which has been increasingly presented through online interactive visualizations (e.g., live dashboards and websites). Thus, it is imperative to understand how older adults interact with and comprehend online COVID-19 interactive visualizations and what challenges they might encounter to make such visualizations more accessible to older adults. We adopted a user-centered approach by inviting older adults to interact with COVID-19 interactive visualizations \rrv{while at the same time verbalizing} their thought processes using a think-aloud protocol. By analyzing their think-aloud verbalizations, we identified four types of thought processes representing how older adults comprehended the visualizations and uncovered the challenges they encountered\st{ with these thought processes}. Furthermore, we also identified the challenges they encountered with seven common types of interaction techniques adopted by the visualizations. Based on the findings, we present design guidelines for making interactive visualizations more accessible to older adults.
\end{abstract}


\input{1-introduction}
\input{2-related-work}

\input{3-1-DesignRationales}
\input{3-2-ThinkAloud-Study}

\input{4-results}
\input{5-discussion}

\input{6-limitations-futurework}
\input{7-conclusion}

\bibliographystyle{apacite}
\bibliography{main.bib}

\section{Biographies}

Mingming Fan is an Assistant Professor in the Computational Media and Arts Thrust and the Division of Integrative Systems and Design at The Hong Kong University of Science and Technology (HKUST) in both Guangzhou and Clear Water Bay campuses. He leads the Accessible \& Pervasive User EXperience (APEX) Group to research at the intersection of Human-Computer Interaction and Accessibility. \\

\noindent{Yiwen Wang is a graduate student in the School of Information at the Rochester Institute of Technology. She has been working with Dr. Mingming Fan on aging and accessibility research.} \\

\noindent{Yuni Xie recently graduated from the School of Information at the Rochester Institute of Technology. She has been working with Dr. Mingming Fan on aging and accessibility research.} \\

\noindent{Franklin Mingzhe Li is a PhD student in the Human-computer interaction institute at Carnegie Mellow University. His research focuses on accessibility.} \\

\noindent{Chunyang Chen is a lecturer (Assistant Professor) in the Faculty of Information Technology, Monash University, Australia. He received his PhD from the School of Computer Science and Engineering, Nanyang Technological University (NTU), Singapore. His research focuses on Software Engineering, Deep Learning and Human-Computer Interaction.} 

\end{document}

%% file: 1-introduction.tex
\section{Introduction}

While people of all age groups have been affected by the COVID-19 pandemic~\citep{world2020coronavirus}, older adults have disproportionately suffered the most severe COVID outcomes~\citep{nanda2020covid,shahid2020covid} and have the highest death rate among all age groups~\citep{Coronavi21:online,Mortalit41:online,COVID19C0:online,COVID19P18:online}.
One critical step to cope with the pandemic is to stay informed about the rapidly evolving situation by acquiring the latest COVID-19 information, such as the latest confirmed cases \rrv{in one's local area}. 
Such real-time COVID-19 information (e.g., the daily new cases) has been widely presented and updated on a daily basis in live dashboards (e.g.,~\citep{WHOCoron51:online,COVID19C75:online,COVID19M32:online,CDCCOVID66:online}) and news media websites (e.g.,~\citep{Interact45:online,Thenovel46:online}) in the form of visualizations.

Visualizations are widely adopted online to convey insights in today's data-driven world. Research suggests that visualizations can help older adults deal with age-related declines~\citep{john1986age,salthouse1996interrelations,strayer1987adult}, such as lowering working memory demands ~\citep{price2016effects}, enhancing prospective memory~\citep{burkard2014implementation}, and facilitating communication~\citep{le2015evaluation,reeder2014assessing,mynatt2000increasing,mynatt2001digital}. However, older adults often encounter difficulties when reading visualizations~\citep{le2015evaluation,bock2016engaging,backonja2016visualization,hakone2016proact}. 
%One important reason causing such difficulties  is the lack of older adults' participation in the design and evaluation of the visualizations that they would use~\citep{backonja2016visualization}. To better understand how older adults have been involved in visualization related research, we reviewed recent ten years' papers (2011-2020) published in three main accessibility and HCI venues (i.e., ASSETS, TACCESS, and CHI) to find the ones containing the following keywords in their title or abstract: ``visualization'', ``visualisation'',  ``visualizing'', ``visualising'', ``visualize'', and ``visualise.'' Among the 146 papers (15 in ASSETS and 131 in CHI) retrieved, 128 included a study with actual users. Within these papers, 75 papers specified the age of their participants. Among these 75 papers, 63 involved young adults. Only 12 papers (8\%) involved older adults.
Such difficulties are further exacerbated by the increasingly popular \textit{interactive} visualizations that require users to actively interact with to reveal hidden information. As COVID-19 visualizations shown in live dashboards and government and new media websites (e.g., ~\citep{WHOCoron51:online,COVID19M32:online,COVID19C75:online,CDCCOVID66:online,Interact45:online,Thenovel46:online}) are often interactive, it is critical to understand whether older adults are able to effectively interact with and comprehend these interactive visualizations and the challenges that they might encounter.
Such an understanding \rrv{could}\st{would} inform the design of interactive visualizations that are more accessible to older adults.

Toward this goal, we adopted a user-centered approach by inviting older adults into the evaluation of COVID-19 interactive visualizations that are critically relevant to their lives during the pandemic. We focused on two research questions (RQs) to explore how older adults interact with COVID-19 interactive visualizations and the challenges that they encounter: 
\begin{itemize}
\item{\textbf{RQ1}: What are older adults' thought processes and the associated challenges when they comprehend COVID-19 interactive visualizations?} 
%Would the three-process visualization comprehension model be applicable?} 
\item{\textbf{RQ2}: What difficulties do older adults encounter with the common \textit{interaction techniques} adopted by the COVID-19 interactive visualizations?}
% \item{\textbf{RQ3}: What are older adults' preferences for interactive visualizations?}
%What are the design implications to make interactive visualizations accessible for older adults?}
\end{itemize}

To answer RQs, we first reviewed COVID-19 related live dashboards and news media websites to identify common types of interactive visualizations and \rrv{curated} a set of representative interactive visualizations. These visualizations covered all common types of interaction techniques adopted in interactive visualizations based on a well-known taxonomy~\citep{yi2007toward}. We further studied the types of data shown in these visualizations and designed a set of tasks for participants to work on in the user study. 
To better understand how they comprehend the visualizations, we asked older adult participants (N=14) to think aloud while working on the tasks \rrv{during the study}. We chose to use the think-aloud protocol because it is \rv{a well-known method for understanding users' thought processes that are otherwise invisible to researchers}~\citep{Thinking18:online} and widely used \rv{in industry to identify user experience problems}~\citep{mcdonald_exploring_2012,fan_practices_2020}.

% After thinking aloud, participants were further interviewed to understand their perceptions of and challenges with the visualizations. 

By analyzing participants' think-aloud verbalizations as well as the corresponding session recordings, we identified five representative thought processes that reflected how older adults comprehended and interacted with the visualizations. 
Moreover, we uncovered the key challenges associated with each type of thought process. Furthermore, we identified the issues that older adults encountered with the seven types of interaction techniques used in visualizations~\citep{yi2007toward}. 
Based on the findings, we derived design guidelines for making interactive visualizations more accessible for older adults. 
% To understand whether the strategies of and difficulties in making sense of the interactive visualizations are unique to older adults, we recruited another X younger adults to participate in the same study. 
% We compared the performance of the two age groups and identified the commonalities and differences in terms of their strategies of and difficulties in making sense of interactive visualizations for the COVID-19 pandemic.  
Finally, we discuss the limitations and future work. In summary, this work makes the following three contributions:
\begin{itemize}
\item {Identification of four types of thought processes that reflect how older adults comprehend COVID-19 interactive visualizations and the associated challenges.}
\item {Identification of the difficulties that older adults encounter with the common interaction techniques adopted by the visualizations.}
% \item{Commonalities and differences between older adults and younger adults in terms of their strategies and difficulties when making sense of interactive visualizations for the pandemic.} 
\item{Design guidelines for making interactive visualizations more accessible for older adults.} 
\end{itemize}

%% file: 2-related-work.tex
\section{Background and Related Work}

Our work was inspired and informed by prior research on \textit{Visualization for Older Adults} and \textit{Visualization Comprehension Theory}.

\subsection{Visualization for Older Adults}
%Ageing is a global challenge facing virtually all countries on earth. 
%Older adults often experience age-related declines related to visual acuity, working memory, prospective memory, and spatial cognition~\citep{john1986age,salthouse1996interrelations,strayer1987adult}.

%Visualization can be a useful tool to better understand older adults' behaviors.
%For example, Franke et al. demonstrated that visualization of older adults' routes on a map can be used with qualitative methods, such as interview, to better understand their mobility~\citep{franke2017grounded}.

Aging is a global challenge, and technologies have been studied to understand challenges faced by older adults and support them to age with dignity. 
Visualization is one such technology and has been explored to understand older adults' behaviors, such as their mobility~\citep{franke2017grounded,o2012visualisation} and face-to-face interactions in the community~\citep{masumoto2017measurement}. 

Visualization has been shown to be beneficial for older adults to deal with many age-related declines.
Visualization can lower older adults' working memory demands. 
%Price et al. examined whether the use of visualization can lower working memory demands in the context of choosing a medicare prescription drug plan~\citep{price2016effects}. 
For example, in the context of choosing a medicare prescription drug plan, Price et al. found that older adults made more accurate decisions in less time when performing a computerized decision-making task to find the best health care plan with the assistance of a simple visualization (i.e., a color-coded table) than without it (i.e., a typical data table)~\citep{price2016effects}.
Similarly, visualization has also been shown to be beneficial for enhancing prospective memory~\citep{burkard2014implementation}.

%mitigate some of the age-related declines. 
Visualization can also facilitate communication between older adults and their healthcare providers~\citep{le2015evaluation,reeder2014assessing} \st{as well as}\rrv{and} promote conversations between them and their family members~\citep{mynatt2000increasing,mynatt2001digital}.
Moreover, visualizations can also help older adults maintain an awareness of \rrv{the progress} of their health goal\st{ progress}~\citep{pham2012effects}, \rv{gain a better understanding of their health conditions~\citep{backonja2016visualization}, and learn potential interaction effects between different medicines~\citep{de2017mevita}}. %\rv{Although previous research suggested simplifying visualization interaction to avoid overwhelming visual information, challenges of utilizing more comprehensive interaction techniques remain unknown.} 

%For example, older adults found value in visualizations to help them engage with their healthcare providers~\citep{le2015evaluation}.
%Visualization is also helpful to stimulate conversation between older adults and their family members. For example, Mynatt et al. designed a digital family portrait that visualizes activities of daily life sensed in a smart home environment as icons bordering the digital picture frame~\citep{mynatt2000increasing,mynatt2001digital}.
%Although such visualization was too complex in conveying ten levels of information, it was found to initiate conversation between older adults and their family members~\citep{mynatt2001digital}.

%Reeder et al. designed sensor visualizations from a half-year pilot study with older adutls and found that visualization of sensor data was useful for older adults when they consulted their health care providers about their activity levels ~\citep{reeder2014assessing}.

Despite these potential benefits of visualizations, few visualizations have been designed \textit{with} and \textit{for} older adults~\citep{backonja2016visualization}. 
The age-related declines among older adults present unique challenges for designing effective visualizations.
Firstly, it is common that older adults would encounter problems when comprehending visualizations~\citep{le2015evaluation,bock2016engaging}.
For example, older adults felt that too much information was presented within the visualization, and they were uncertain about what elements of the visualization to focus on~\citep{le2015evaluation} or \rv{had difficulty interpreting \rrv{abstract} visualizations\st{ that were abstract}~\citep{backonja2016visualization}.}
Secondly, not all visualizations are equally effective for older adults. 
For example, Hakone et al. found that pie charts \rrv{sampling along}\st{that
sampled} temporal dimensions were more effective at communicating change-over-time data when compared to temporal
area charts among older patients~\citep{hakone2016proact}.
Consequently, without taking older adults' experiences into consideration, the increasingly pervasive visualizations might become inaccessible to them. 

One key step to making visualizations more accessible \rrv{to}\st{for} older adults is to understand how they comprehend visualizations.
%Recently, researchers began to explore this area. 
For example, Ahmed et al. conducted a participatory design with older\st{ adult} patients to design visualizations for data generated from cardiac re-synchronization therapy devices to increase their engagement in care for a patient-facing dashboard~\citep{ahmed2019visualization}. Similarly, Le et al. conducted a participatory design with older adults~\citep{le2014design} to design health visualizations and later interviewed older adults to understand how they used health visualizations and potential barriers~\citep{le2018understanding}. 
They found that older adults felt that contextual information about visualizations was helpful and that their computer literacy might have affected their understanding.

% Le et al. conducted focus groups with older adults to understand their usage and preferences of three types visual encodings of health data: the bar graph, the radial plot, and the light ball metaphor visualization.~\citep{le2015evaluation}.
% %They found that older adults applied a high-level approach toward processing visual information by focusing on holistic scores of wellness first and then, if necessary, examining components of wellness. 
% They found that older adults felt that too much information was presented within the visualization and were uncertain on what elements of the visualization to focus and the abstractions of data led to future confusion.  

% They provided specific recommendations to improve the three visualizations used in the study.
% One key takeaway is that having multiple cues to reinforce a visual trend detracts from cognitive efficiency and thus visual cues should be used in moderation. 

% Researchers have begun to understand how older adults use visualizations.
% For example, Le et al. conducted interview studies with 21 older adults to understand how older adults use health visualizations and potential barriers that impact utility~\citep{le2018understanding}. 
% They found that older adults found visualizations were useful in detecting trends and wished to have more contextual information to better understand visualization.
% Older adults expressed concerns with not be able to access or manipulate the visualizations due to a lack of understanding in using computers.
Inspired by this line of research that primarily focused on non-interactive visualizations for older adults~\citep{ahmed2019visualization,le2014design,le2018understanding,le2015evaluation}, we seek to understand how older adult use and perceive \textit{interactive visualizations}. \rrv{In this work, we focused on interactive visualizations} related to COVID-19, which is critically relevant to older adults' lives.
%on the web, which have been increasingly prevalent partially due to open-source visualization libraries and languages (e.g., D3~\citep{bostock2011d3}, Vega~\citep{satyanarayan2015reactive}, Vega-lite~\citep{satyanarayan2016vega}, and Lyra~\citep{satyanarayan2014lyra}) that have made the development of interactive visualizations easier than ever before. 
To do so, we adopt \textit{usability testing with the think-aloud protocol} to better understand older adults' thought processes, that are otherwise unavailable to researchers, when they comprehend interactive visualizations.

\subsection{Visualization Comprehension Theory}
\label{sec:VisualizationComprehension}
%\subsection{How do people comprehend visualizations?}
% cognitive principles of visualization 

% Cleveland and McGill proposed the scientific foundation of graphical methods for data analysis and for data presentation~\citep{cleveland1984graphical}.

% \textit{graphical perception} refers to the visual decoding of information encoded on graphs. 
% The first part is an identification of a set of \textit{elementary perceptual tasks} that are carried out when people extract quantitative information from graphs. 
% The second part is an ordering of the tasks on the basis of how accurately people perform them. 

% They tested the theory with bar charts, divided bar charts, pie charts, and statistical maps with shading. 

Building on top of earlier work \rrv{investigating}\st{on} \rrv{human} visualization comprehension \rrv{processes} in 1980s ad 1990s (e.g.,~\cite{bertin1983semiology,pinker1990theory,cleveland1985graphical}), Carpenter and Shah proposed a \textbf{three-process visualization comprehension theory} to explain how people comprehend visualizations: \textit{encoding visual information}; \textit{relating visual features to concepts}; \textit{associating concepts with existing knowledge}~\citep{shah2002review}. Further, their experimental results showed that visualization comprehension often involves multiple, integrated, and iterative cycles of these three processes~\citep{carpenter1998model}.

\textit{Encoding visual information} involves identifying visual characteristics of a visualization, which influence\st{s} how effective a viewer can encode graphical information~\citep{bertin1983semiology}.
Bertin characterized visual variables for building graphs: position, size, shape, value, color, orientation, and texture~\citep{bertin1983semiology}. 
Cleveland and McGill identified six similar elementary perceptual tasks that people carry out when extracting quantitative information from graphs~\citep{cleveland1984graphical} and conducted a study to rank people's judgments on these six tasks from the most accurate to the least accurate: position along a common scale, position along a nonaligned scale, length and angle, area, volume, shading~\citep{cleveland1986experiment}.

\textit{Relating visual features to concepts} is the translation of visual features into the conceptual relations represented by those features~\citep{pinker1990theory,Kosslyn89}.
Pinker proposed a set of cognitive operations that are executed when processing a graph to create a mapping that relates visual features to conceptual relations found in the display~\citep{pinker1990theory}.
%No experimentation accompanies Pinker's modeling. 
Specifically, visualization is mapped into conceptual relations such as differences in size, changes in trend, and differences in spatial location~\citep{Kosslyn89,bertin2011graphics}. 
Thus, this process involves interpreting visual pattern quantitatively, such as an upwardly curved line means an accelerating trend~\citep{carpenter1998model}. 
Further, the translation process is complicated by multiple graphical features, such as multiple lines in a graph~\citep{carpenter1998model}.
Thus, errors in interpretation may occur when visual characteristics do not effectively get translated into concepts or relationships~\citep{Kosslyn89,larkin1987diagram,pinker1990theory}.
Moreover, the relative ease of mappings between visual features and referents in visualizations can also affect this translation process. For example, horizontally oriented bars may be better for representing variables for which the horizontal dimension is more meaningful than the vertical dimension, such as in depicting data about distance traveled~\citep{kosslyn1994elements}. 

\textit{Associating concepts with existing knowledge} refers to the process of leveraging existing knowledge in the interpretation of visualizations. Existing knowledge may provide contexts \st{for comprehension of }\rrv{to comprehend} conceptual elements~\citep{bertin1983semiology}. Shah found that students' existing knowledge and expectations of content affected their interpretation\rrv{s} of graphs~\citep{shah1999role,shah1995conceptual}. A viewer may make an error in the interpretation of a visualization when its visual feature does not automatically evoke a particular fact or relationship for her (e.g., ~\cite{shah1999role,stenning1995cognitive,pinker1990theory, larkin1987diagram}).
Le et al. suggested that the knowledge gap in health information among older adults may contribute to the challenges that they encounter in comprehending health-related visualization\rrv{s}~\citep{le2015evaluation,le2014thesis}.

% existing knowledge provides context for conceptual elements when users interpret the visualizations~\citep{bertin1983semiology}. 

% From a design perspective, this suggests the need to articular the underlying structure of the visualizations and to relate it to familiar existing experiences of users~\citep{le2015evaluation}.

% The value of the visualization is impacted by the match between structure and function~\citep{larkin1987diagram}.

% The three-process model of visualization comprehension has been well studied and validated. 
% Thus, we also apply this model when we analyze how older adults comprehend visualizations.
The three-process theory of visualization comprehension was developed and validated with mostly \textit{static} visualizations on printed paper or computer screen in the 1980s and 1990s. 
In the recent decade, \textit{interactive} visualizations have become increasingly popular on the web, such as in news media reports and dashboards.
Compared to static visualizations, interactive visualizations allow users to explore data \rrv{dynamically} from different perspectives through interactions.
As interactive visualizations are often not designed and evaluated with older adults, they may encounter difficulties in comprehending and interacting with such interactive visualizations. 
Thus, it is important to examine whether the three-process theory is still applicable by studying how \textit{older adults} comprehend \textit{interactive} visualizations and to identify ways to improve such interactive visualizations for older adults by uncovering potential challenges \st{older adults}\rrv{they} may encounter.

%% file: 3-1-DesignRationales.tex
\section{Method}
% Our goal is to understand strategies and difficulties that older adult would have when making sense of the interactive visualizations related to COVID-19. 

To answer RQs, we conducted online think-aloud usability testing with older adults. 
During think-aloud usability testing, older adults worked on tasks related to the COVID-19 interactive visualizations while at the same time verbalizing their thought processes. 
Compared to other approaches such as survey or interview, this approach allowed us to gain insights into older adults' \rrv{invisible}\st{hidden} thinking processes while they were comprehending interactive visualizations.
% Furthermore, after interacting with the visualizations, older adults would have more concrete experiences to share during the follow-up interviews. 

% While curating the set of interactive visualizations, we picked a timely theme that was relevant and important for older adults to seek information for---COVID-19.
% Compared to other age groups, older adults have higher chances of contracting the novel coronavirus and also much higher death rates~\citep{COVID19P18:online}.
% Thus, it is important for older adults to quickly grasp the latest information about COVID so that they could plan their daily activities, such as grocery shopping, in an informed manner.
% As major news media reports and live dashboards release daily updates about COVID situations via interactive visualizations, such as WHO~\citep{WHOCoron51:online}, %it is important for older adults to comprehend such interactive visualizations. 
% We sought out to curate a dataset of representative interactive visualizations for older adults to work on in the online think-aloud usability testing. 

We followed three design considerations while curating the dataset of interactive COVID-19 visualizations and corresponding tasks for the study. 
Firstly, we reviewed commonly appearing interactive visualizations in major news media websites and live dashboards about COVID-19 to ensure that our dataset covered all common types of visualizations. 
Secondly, we consulted the visualization literature to understand the interaction techniques widely adopted in interactive visualizations and ensured that \st{these typical}\rrv{common} interaction techniques were well represented in our dataset.
Lastly, we studied the information conveyed in these COVID-19 interactive visualizations and designed the corresponding tasks based on the information.  
In the next three subsections, we explain \st{the details of }these three steps \rrv{in detail}. 

\begin{figure*}[h]
  \centering
  \includegraphics[width=\linewidth]{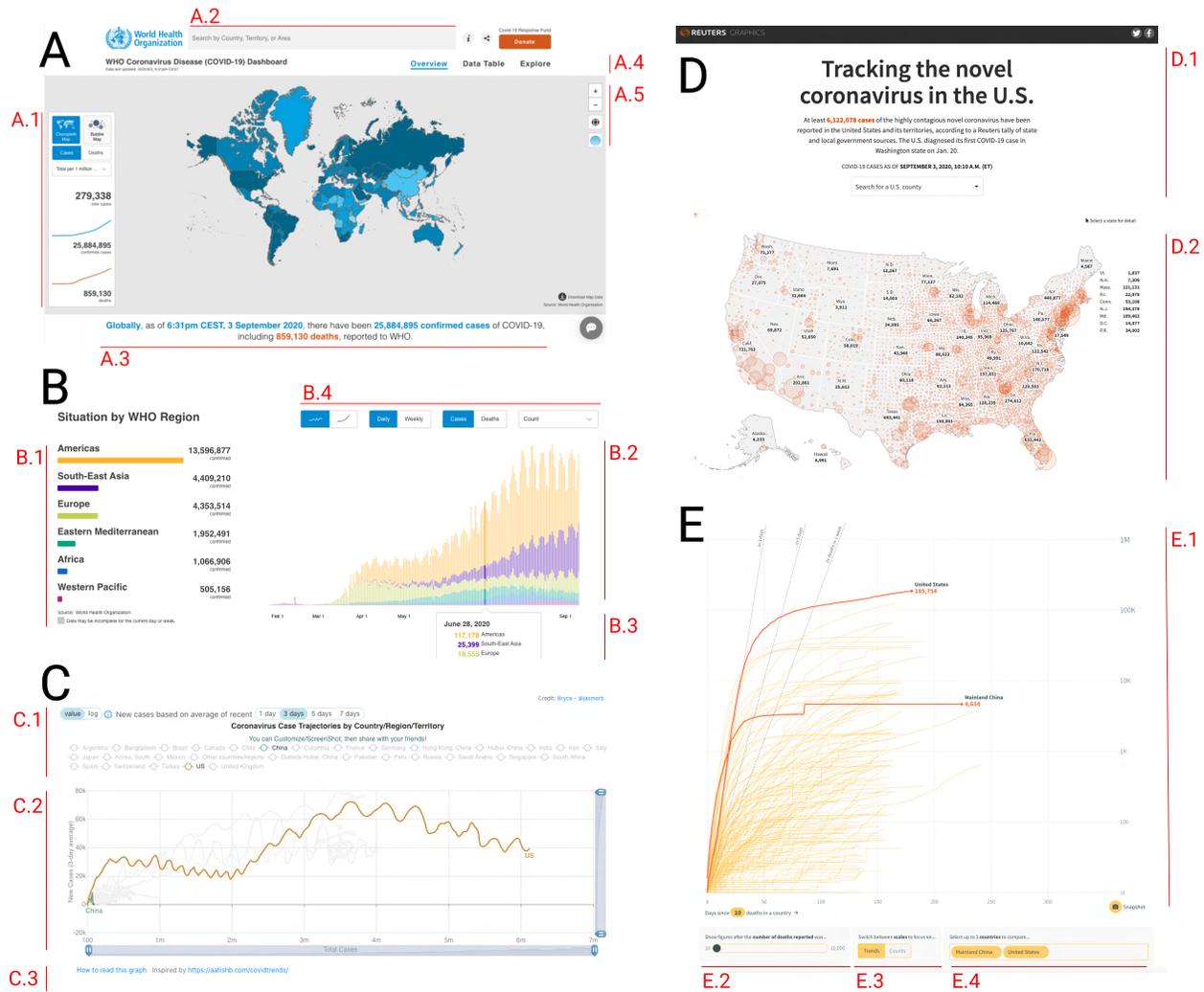}
  \caption{The five interactive visualizations (labeled A to E) used in the study}
%   \Description{The figure falls into five parts. WHO Figure A map includes the world map in the center colored by different blue level and several features in the surrounding area. WHO Figure B bar chart has one horizontal bar chart in the left described the total confirmed cases of regions and a stacked bar chart in the right, 1point 3 Figure C line chart presented COVID-19 cases by country/Region/Territory, x-axis is the total cases, and y-axis is new cases in 3-day average. Reuters Figure D map shows the U.S COVID-19 cases by states and regions, bubbles, and colors are used in the map. Reuters Figure E line chart presents the death cases by countries, X-axis is days since N deaths in a country and Y-axis is the number of death cases}
  \label{fig:interactivevisualizations}
\end{figure*}

% To answer the RQs, we would first need to curate a list of representative interactive visualizations and the corresponding tasks.
% To do so, we reviewed COVID-19 related websites to identify the the interactive visualizations.
% We further analyzed information presented in these visualizations and identified the common types of interactions in interactive visualizations from the literature with the aim of deriving a list of tasks that would be representative of the COVID-19 interactive visualizations as well as cover common types of interactions in the literature. 
% The list of interactive visualizations and the tasks would be used in our user study.
% Next, we present the details of design rationales for selecting the interactive visualizations and the list of tasks.   

\subsection{Interactive Visualizations for the Study}
% \ming{TODO: update the table to show all the visualizations we used; what does the number 
% 172 refer to? - 57 news and dashboards,172 visualizations, 16 sources in total}

% Line Chart	Combo Chart	Choropleth Map	Bubble Map Chart	Bar Chart	Timeline	Stacked Chart	100% Stacked Chart	Area chart	Pie Chart	Doughnut Chart	Bubble chart	Table	Infographic chart	Scatter plot
% 31	14	24	13	23	4	4	4	13	1	1	2	29	4	4

We searched \rv{the following} keywords in Google: \textit{COVID, coronavirus,} \rv{\textit{COVID-19, covid, case, data, interactive, visualization, chart, map, and graphic}}. From the top \rv{10 pages of the search results up until June 30, 2020.} we \rv{identified} 57 dashboards and news reports about COVID \rv{from 16 well known organizations}. 
The number of dashboards or news reports from these 16 sources was as follows: Word Health Organization (WHO) dashboard (1), 1point3acres dashboard (1), Johns Hopkins University (JHU) dashboard (1), The New York Times (10), Reuters (7), Bloomberg (6), CNN (5), Forbes (5), NBC (5), The Economist (5), The Washington Post (3), BBC (2), NPR (2), The World meters (2), ABC (1), and state government COVID website (1).  

%\st{We further studied 172 COVID visualizations within these 57 dashboards and news reports and finally selected five visualizations to be used in our study. These five visualizations were selected because they covered all common types of interaction types in visualizations, which will be explained in the next Section (Section~\ref{InteractionTypes}).}

\rv{From these 57 dashboards and news reports, the first three authors identified 172 COVID-related visualizations. They analyzed the visualization types (e.g., line graph, bar chart) and interaction techniques (e.g., zoom, filter) used in these visualizations and chose five examples for the user study.} Figure~\ref{fig:interactivevisualizations} shows the five interactive visualizations, which are labeled from A to E. \rv{These five example visualizations were chosen to cover the common visualization types and interaction techniques; the exact number of visualizations \rv{chosen was} determined after two rounds of pilot studies to ensure that} the duration of the study was around an hour. The interaction techniques used in these visualizations will be elaborated on in the next section.

% \begin{table*}[tbh]
%   \caption{The the types of interactive visualizations and their frequencies}
%   \label{tab:interactivevisuaizations}
%   \begin{tabular}{p{3cm}|p{3cm}|p{4cm}|p{6cm}}
%     \toprule
%     X & Visualization types & Frequency & Examples\\
%     \midrule
%     A & Map  & 37 (22\%) & \textit{website}\\
%     \hline
%     B & Line  & 31 (18\%) & \textit{website}\\
%     \hline
%     C & Bar  & 23 (13\%) & \textit{website}\\
%     \bottomrule
% \end{tabular}
% \end{table*}

% Table~\ref{tab:interactivevisuaizations} shows the types of interactive visualizations, their frequencies, and examples. 

\begin{figure*}[hbt!]
  \centering
  \includegraphics[width=0.98\linewidth]{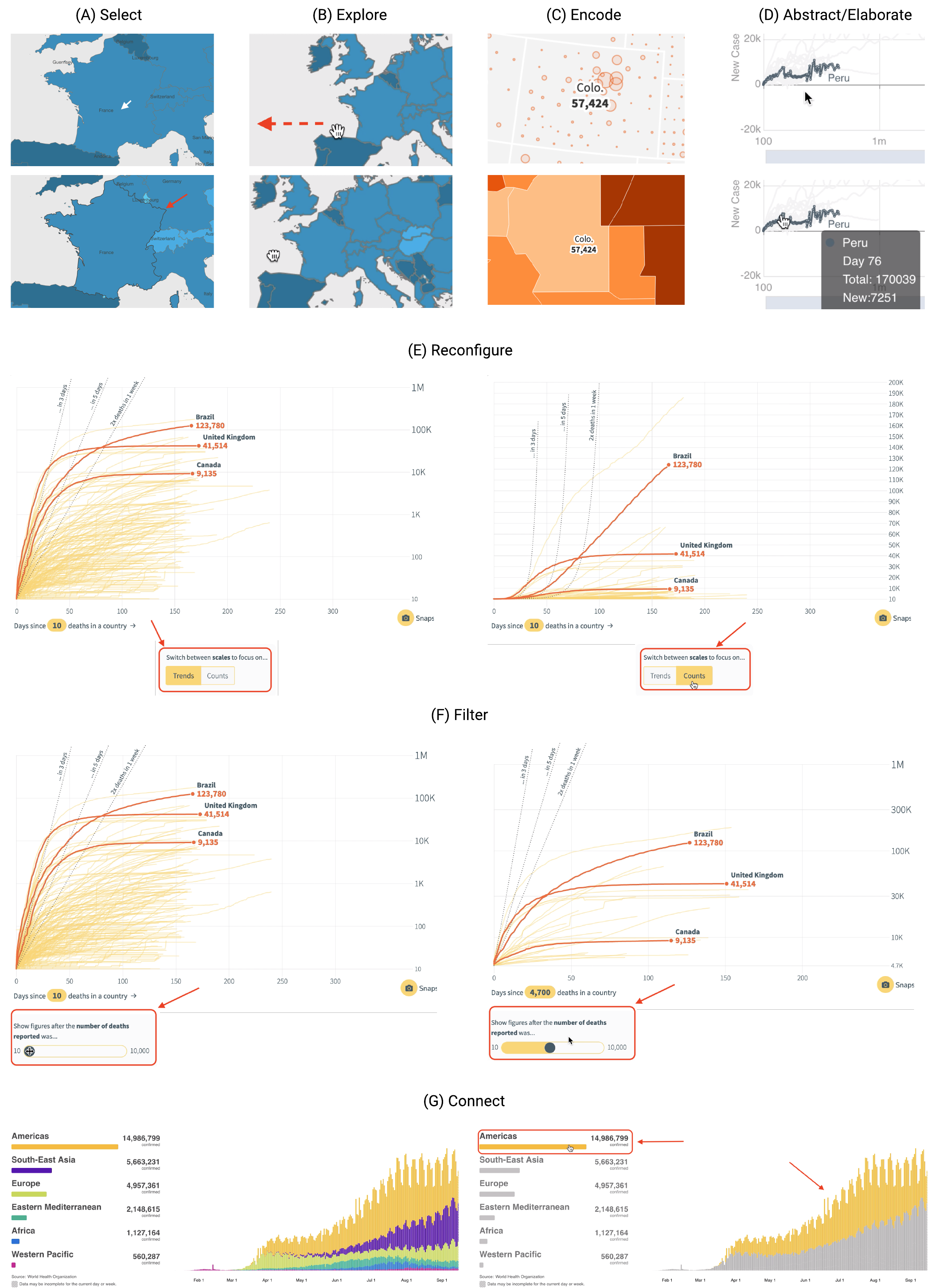}
  \caption{The instantiation of each interaction technique~\citep{yi2007toward} in example visualizations in Figure~\ref{fig:interactivevisualizations}. The appendix includes all the relevant interaction techniques that each visualization \rrv{in Figure~\ref{fig:interactivevisualizations}} instantiates.}
%   \Description{This figure falls into eight parts. Select A shows the outline of France has been highlighted by black color. Explore B shows moving the world map by cursor. Reconfigure C is changing the line chart between trend and count. Encode D is the US map encodes the cases of states from bubble map to choropleth map. Abstract/Elaborate E shows hovering over the lines to pop up further information. Filter F is a scrollbar below the line chart, which can filter the data in the graph. Connect G is by hovering the horizontal bar chart in the left, connecting with the data colored in the stacked bar chart in the right}
  \label{fig:interactiontype}
\end{figure*}
\subsection{Interaction Techniques in the Visualizations}
\label{InteractionTypes}

%(COVID Dashboard World Overview)
%(COVID Dashboard Regional Situations)
%(COVID Dashboard)
% (COVID Tracker for the US)
% (Countries' COVID Trends)
%~\footnote{https://covid19.who.int/}   
%~\footnote{https://coronavirus.1point3acres.com/en}
%~\footnote{https://graphics.reuters.com/HEALTH-CORONAVIRUS-USA/0100B5K8423/index.html}
%~\footnote{https://graphics.reuters.com/HEALTH-CORONAVIRUS/yxmvjookdpr/index.html}
\rrv{The} visualization community has proposed taxonomies to characterize interaction techniques used in visualizations~\citep{dimara2019interaction,yi2007toward}. We compared two primary taxonomies~\citep{dimara2019interaction,yi2007toward} and\st{applied them to our visualizations. Finally, we} chose to use Yi et al.'s taxonomy~\citep{yi2007toward} because this taxonomy characterized the interaction techniques in our visualizations better and was also more widely used. Yi et al.'s taxonomy includes seven types of interaction techniques commonly used in visualizations: \textit{Select}, \textit{Explore}, \textit{Reconfigure}, \textit{Encode}, \textit{Abstract/Elaborate}, \textit{Filter}, and \textit{Connect}.~\citep{yi2007toward}.
%, and \textit{Others}
%The taxonomy has been widely cited and used in the literature. 
%As a result, we also adopted this taxonomy and ensured that our curated dataset of interactive visualizations covered all these types of interactions.

To illustrate how each of the seven \st{type of }interaction techniques is instantiated, Figure~\ref{fig:interactiontype} shows one example drawn from our visualizations in Figure~\ref{fig:interactivevisualizations}. Next, we explain how these interaction techniques are instantiated in details. We have included more details to explain how each visualization instantiates all the relevant interaction techniques in the supplemental file.

\textbf{Select}: Select interaction techniques allow users to mark a data item(s) of interest visually distinctive so that they could easily keep track of it when there are too many data items on a view or when representations of data are changed. Figure~\ref{fig:interactiontype}A shows an example \st{of a }Selection technique. 

% 1. Select: mark something as interesting
% Select interaction techniques provide users with the ability to mark a data item(s) of interest to keep track of it. When too many data items are presented on a view, or when representations are changed, it is difficult for users to follow items of interest. By making items of interest visually distinctive, users can easily keep track of them even in a large data set and/or with changes in representations
 
\textbf{Explore}: Exploration interaction techniques enable users to examine a different subset of a data set. Due to some combination of the large scale of the data set, view, and/or limited screen space, users often only see a limited number of data items at a time. Thus, Exploration interaction techniques allow users to gain understanding and insights from a subset of the data before moving to view different subsets of the data. Figure~\ref{fig:interactiontype}B shows an example. The panning interaction on this map-based visualizations allows users to explore different parts of the data set one at a time. 

% show me something else (Moving the map scene with mouse)
% Explore interaction techniques enable users to examine a different subset of data cases. When users view data using an Infovis system, they often can only see a limited number of data items at a time because of some combination of the large scale of the data set, view and/or screen limitations, and fundamental perceptual and cognitive limitations in human information processing. Explore interactions do not necessarily make complete changes in the data being viewed, however. More frequently, some new data items enter the view as others are removed. 1) The most common Explore interaction technique in our survey is panning. Panning is often achieved by a special mode where the user grabs the scene and moves it with a mouse or by simply altering the view via scrollbars. 2) Another example of an Explore interaction is the Direct-Walk technique. Direct-Walk allows users to smoothly move the viewing focus from one position in information structure to another by “a series of mouse points or other direct-manipulation methods”

\textbf{Encode}: Encode interaction techniques enable users to alter the visual representation of the data, including the visual appearance (e.g., color, size, and shape) of data items. 
Visual representations are important because they affect pre-attentive cognition and are related to how users understand relationships and distributions of the data items.

Changing how the data is represented, such as from a bar chart to a map, is an example of Encode. 
Another example of Encode is to alter the color encoding of a data set.
Similarly, altering size, orientation, font, and shape encodings are also examples of Encode. 
Figure~\ref{fig:interactiontype}C shows an example \st{of }Encode \rrv{interaction technique}\st{in our data set}. 

% 4. Encode: show me a different representation (User can change color, size, font, and shape)
% Encode techniques enable users to alter the fundamental visual representation of the data including visual appearance (e.g., color, size, and shape) of each data element. Simply changing how the data is represented (e.g., changing a pie chart to a histogram) is an example of Encode. By changing a type of representation, users expect to uncover new aspects of relationships. Another widely used technique of Encode is the set of interaction techniques that alter the color encoding of a data set. Enable users to adjust a color or a spectrum of colors for a certain variable. Beyond color encoding, many systems provide other encoding techniques, such as size (e.g., Dust & Magnet [55]), orientation (e.g., Polaris [43]), font (e.g., SemaSpace [31]), and shape (e.g., Spotfire [2]).

\textbf{Abstract/Elaborate}: Abstract/Elaborate interaction techniques allow users to adjust the level of abstraction of data representation.
One example is details-on-demand techniques, such as the tool-tip interaction, which provide\st{s} more details when a mouse cursor hovers over a data item.
Another example is geometric zooming. Geometirc zooming allows users to change the scale of \st{representation}\rrv{represented data} so that they could see an overview of a larger data set (with zoom-out) and a detailed view of a smaller data set (with zoom-in). The representation of underline data is not altered during geometric zoom. Figure~\ref{fig:interactiontype}D shows examples of Abstract/Elaborate in our data set.

% 5. Abstract/Elaborate: show me more or less detail (Tooltip: Hover over to see more detail; Zooming: the representation is not fundamentally altered during zooming)
% Abstract/Elaborate interaction techniques provide users with the ability to adjust the level of abstraction of a data representation. These types of interactions allow users to alter the representation from an overview down to details of individual data cases and often many levels in-between. An exemplary interaction technique in this category is any technique from the set of details-on-demand operations. For example, the drill-down operation in a treemap visualization, such as SequoiaView (formerly known Cushion Tree [48]), allows a user to examine a particular sub-tree within an information hierarchy. Furthermore, simple tool-tip interaction techniques that provide detailed information when a mouse cursor hovers over a data item also belong to this category: for example, SeeIT [1] as shown in Fig. 6.
% Another very common but slightly complex example of Abstract/Elaborate techniques is zooming (or geometric zooming if it is to be distinguished from semantic zooming). Through zooming, users can simply change the scale of a representation so that they can see an overview of a larger data set (using zoom-out) or the detailed view of a smaller data set (using zoom-in). A key point here is that the representation is not fundamentally altered during zooming. Details simply come more clearly into focus or fade away into context.

\textbf{Reconfigure}: Reconfigure interaction techniques allow for changing the spatial arrangement of data representations (e.g., how data items are arranged or \rrv{aligned}\st{the alignment of data items}) to have different perspectives onto the data set.

Sorting and rearranging columns \st{operations }are examples of Reconfigure interaction techniques. Further, changing the attributes assigned to x- and y-axes is another example, which \st{would }change\rrv{s} the relationships between the data items visualized and thus provide\\rc{s} different perspectives. What distinguishes ``Reconfigure'' from ``Abstract/Elaborate'' is that the representation is changed in ``Reconfigure''.
Figure~\ref{fig:interactiontype}E shows an example \rrv{in}\st{from} our data set. 

% 3. Reconfigure: show me a different arrangement - the relationships between data items cannot be changed (Sorting and rearranging columns)
% Reconfigure interaction techniques provide users with different perspectives onto the data set by changing the spatial arrangement of representations. One of the essential purposes of Infovis is to reveal hidden characteristics of data and the relationships between them. The sorting and rearranging columns operations in TableLens [33] are good examples of Reconfigure techniques.
% -   	Not belongs to reconfigure: Changing the attributes assigned to x- and y-axes changes the sets of attributes or variables to be examined among the entire data set, so it eventually changes relationships between data items and provides different perspectives.
% Other system’s interaction techniques allow users to move data items more freely to make the arrangement more suitable for their mental model.
% Reconfigure techniques also include a set of interaction techniques reducing occlusions.

\textbf{Filter}: Filter interaction techniques allow users to change the set of data items being visualized based on some specific conditions. 
When using this type of interaction, users specify a range or condition so that only data items meeting the criteria are shown. 
Because the actual data items are usually unchanged, the view would resume when users remove the filtering criteria.
In a similar vein, filter interaction techniques do not change the perspective on the data but just specific conditions for data to be shown.
This category includes interaction techniques that allow users to \st{specific}\rrv{specify} ranges, keywords, or queries. 
Figure~\ref{fig:interactiontype}F shows an example \rrv{in}\st{from} our data set. 
 
% 6. Filter: show me something conditionally (Users can specify a range or condition)
% Filter interaction techniques enable users to change the set of data items being presented based on some specific conditions. In this type of interaction, users specify a range or condition, so that only data items meeting those criteria are presented. Data items outside of the range or not satisfying the condition are hidden from the display or shown differently, but the actual data usually remain unchanged so that whenever users reset the criteria, the hidden or differently shown data items can be recovered. The user is not changing perspective on the data, just specifying conditions on which data are shown. The Attribute Explorer [39] extends dynamic query capabilities by changing the colors of filtered data items rather than removing them from the display, as shown in Fig. 7. QuerySketch [52] is yet another interesting example of Filter technique. QuerySketch allows users to draw a line graph freehand, and then the system retrieves and presents data cases with similar graphs. These graphs frequently represent time series data.
 
\textbf{Connect}: Connect interaction techniques highlight associations and relationships between data items. 
When more than one view is used to show different representations of the same data set, it can be difficult to find the corresponding item of a data point in a different view. 
To alleviate such difficulty, Connect interaction techniques often highlight the corresponding item in a different view when the user selects an item in one view.  
Similarly, Connection interactions also apply to the same view by highlighting the related items to a selected data point. 
Further, Connect interactions can also reveal data items that are originally hidden in the view. 
Figure~\ref{fig:interactiontype}G shows an example in our data set. When hovering over a horizontal bar (e.g., Americas) in Figure~\ref{fig:interactivevisualizations}B.1, the corresponding part of the stacked bar chart in  Figure~\ref{fig:interactivevisualizations}B.2 is highlighted.

Table~\ref{tab:tasks} shows how these interaction techniques are instantiated in each of the visualizations in Figure~\ref{fig:interactivevisualizations}.

% \begin{table*}[h]
%   \caption{The types of Interaction techniques instantiated in each of the five visualizations in Figure~\ref{fig:interactivevisualizations}.}
%   %\ming{E also has Select; Connect is between A and B; I've updated the table.}
%   \label{tab:tasks}
% %   \begin{tabular}{p{0.4cm}|p{3.8cm}|p{0.8cm}|p{1.0cm}|p{1.1cm}|p{1.5cm}|p{1.1cm}|p{2.4cm}|p{1.0cm}|p{1.1cm}}
%   \begin{tabular}{p{0.4cm}|p{2.0cm}|p{0.6cm}|p{0.7cm}|p{1.0cm}|p{1.5cm}|p{1.0cm}|p{2.4cm}|p{0.9cm}|p{1.1cm}}
%     \toprule
% 	& Source & Type & Select & Explore & Reconfigure & Encode & Abstract/Elaborate & Filter & Connect\\
%     \hline
% A	&	WHO 1~\citep{WHOCoron17:online} & Map & \checkmark	& \checkmark	& \checkmark	&\checkmark	&\checkmark	&\checkmark	& \checkmark \\
% \hline
% B	&	WHO 2~\citep{WHOCoron17:online} & Bar &	& 	& \checkmark	&\checkmark	&\checkmark	&	& \checkmark \\
% \hline
% C	& 1Point3Acres~\citep{COVID19C75:online} & Line & \checkmark	& 	& \checkmark	&	&\checkmark	&	&  \\
% \hline
% D	&	Reuters 1~\citep{Thenovel46:online}  & Map & 	& \checkmark	& 	&\checkmark	&\checkmark	&\checkmark	&  \\
% \hline
% E	&	Reuters 2~\citep{Interact45:online} & Line & \checkmark	& 	& \checkmark	&	& \checkmark	&\checkmark	&  \\
%     \midrule
%     %\bottomrule
% \end{tabular}
% \end{table*}

\begin{table*}[h]
  \caption{The types of Interaction techniques instantiated in each of the five visualizations in Figure~\ref{fig:interactivevisualizations}. The visualizations collectively cover all the interaction techniques.}
  %\ming{E also has Select; Connect is between A and B; I've updated the table.}
  \label{tab:tasks}
%   \begin{tabular}{p{0.4cm}|p{3.8cm}|p{0.8cm}|p{1.0cm}|p{1.1cm}|p{1.5cm}|p{1.1cm}|p{2.4cm}|p{1.0cm}|p{1.1cm}}
  \begin{tabular}{p{1.2cm}|p{2.0cm}|p{0.6cm}|p{0.7cm}|p{1.0cm}|p{1.6cm}|p{1.0cm}|p{1.5cm}|p{0.8cm}|p{0.9cm}}
    \toprule
Vis. IDs	& Source & Type &  Select & Explore & Reconfigure & Encode & Abstract / Elaborate & Filter & Connect\\
    \hline
A	&	WHO 1~\footnotemark[1] & Map & \checkmark	& \checkmark	& \checkmark	&\checkmark	&\checkmark	&\checkmark	& \checkmark \\
\hline
B	&	WHO 2~\footnotemark[1] & Bar  &	& 	& \checkmark	&\checkmark	&\checkmark	&	& \checkmark \\
\hline
C	& 1Point3Acres~\footnotemark[2] & Line  & \checkmark	& 	& \checkmark	&	&\checkmark	&	&  \\
\hline
D	&	Reuters 1~\footnote[3]  & Map  & 	& \checkmark	& 	&\checkmark	&\checkmark	&\checkmark	&  \\
\hline
E	&	Reuters 2~\footnote[4] & Line & \checkmark	& 	& \checkmark	&	& \checkmark	&\checkmark	&  \\
    \midrule
    %\bottomrule
\end{tabular}
\end{table*}

\subsection{Tasks for the Visualizations}
We had three design considerations when designing the tasks for the visualizations that participants would comprehend during the think-aloud usability testing. 
First, the tasks should \rv{utilize} the information shown in the visualizations. 
To do this, \rv{the first three authors independently} analyzed the captions and the labels of the axes of the COVID-19 visualizations in our data set and extracted terms and phrases from the captions and the labels to construct a taxonomy of COVID-19 information. \rv{They further discussed their results to gain a consensus of the key terms and phrases}. Then, they applied an affinity diagramming approach on these terms and phrases to derive two themes related to COVID-19: \textit{expressions} and \textit{impacts}. 
Within the COVID-19 expressions theme, there were five sub-themes: \textit{Cases} (cases by region,population,day,condition,people, active cases, recovered cases); \textit{Deaths} (deaths by region,day,people, estimated deaths, other deaths); \textit{Rates} (fatality rate, Change frequency, growth rate); \textit{Tests} (tests by day, region, people); and \textit{Others} (population, time).
Within the COVID-19 impacts theme, there were four sub-themes: \textit{Health} (symptoms, mental,psychological health, other diseases), \textit{Behaviors} (activities, spreading index), \textit{Resources} (hospitals, transportation), and \textit{Economy and Policies} (finance, government policies).

\footnotetext[1]{\url{https://covid19.who.int/}}
\footnotetext[2]{\url{https://coronavirus.1point3acres.com/en}}
\footnotetext[3]{\url{https://graphics.reuters.com/HEALTH-CORONAVIRUS-USA/0100B5K8423/index.html}}
\footnotetext[4]{\url{https://graphics.reuters.com/HEALTH-CORONAVIRUS/yxmvjookdpr/index.html}}

\begin{table*}[h]
  \caption{The tasks for the visualizations in Figure~\ref{fig:interactivevisualizations}.}
  \label{tab:tasks-for-visualizations}
  %\resizebox{\linewidth}{!}{
  %{p{1.4cm}|p{5.5cm}|p{7.2cm}}
  \begin{tabular}{p{1.3cm}|p{13.2cm}}
    \toprule
   Vis. IDs & Tasks\\
    \midrule
    A & find out the confirmed COVID cases “per 1 million population” in France \\
    \hline
    B & find out the daily confirmed COVID cases on June 28 in Americas; \newline find out the date when Americas exceeded Europe in terms of the cumulative COVID cases \\
    \hline
    C & find out which country the light-color line that is above the line representing the US represents \\
    \hline  
    D & find out the county that has the lowest positive COVID-19 cases in Nevada\\
    \hline
    E & find out which country has the least number of COVID cases: Brazil, Canada, and Germany; \newline find out which countries have over 10, 000 death cases for more than 100 days \\
  \bottomrule
\end{tabular}
%}
\end{table*}

% 1. COVID-19 Expression
%     - Cases(Cases by region, Cases by population,m Cases by day, Cases by condition, Active cases, Recovered cases, Cases by people) 
%     - Deaths(Deaths by region, Deaths by day, Deaths by people, Estimated deaths, Other deaths)
%     - Change rate(Fatality rate, Change frequency, Growth rate)
%     - Test cases(Tests by day, Tests by region, Tests by people)
%     - Other variable(Population, Time)
% 2. COVID-19 Impacts
%     - People's behavior (Human activity, spreading index) 
%     - Health related information (Other diseases, Symptoms development, Psychological)
%     - Social Resources (Hospitalized, Transportation)
%     - Policy&Economic (Political trends, Fiance, Reaction)

Second, the tasks should cover all common interaction types in Section~\ref{InteractionTypes}.
Third, the tasks should be of different levels of complexity. 
%Table~\ref{} shows the final tasks. 
Following these design considerations, we designed the tasks for the visualizations in Table~\ref{tab:tasks-for-visualizations}. %\rv{We also consider these tasks are typical for older adults to find the information they want to know when they look into the visualizations (e.g., find the daily confirmed COVID cases.). Basically, our tasks include the key functions of the visualizations (e.g., zoom-in, filter).} 
Note that the tasks were different for each visualization.  

%% file: 3-2-ThinkAloud-Study.tex
\section{User Study}
%We present the details about our IRB-approved user study.
\subsection{Participants}
% \ming{TODO: demographics information of the participants; and how we recruited them.}
This research was approved by IRB at our institution. We recruited participants through advertisements posted on social media platforms, word-of-mouth, and snowball sampling. 
We first conducted a pilot study with two older adults (one female and one male, aged 64 and 77) and used the data to revise the visualizations and the corresponding tasks and to ensure the study procedure was appropriate. 
We then conducted the formal online think-aloud usability testing with eighteen older adults (eight females and ten males). Participants lived in Canada ($N=15$) and USA ($N=3$). Their ages were between 60 and 78 ($M=67, SD=5, median = 65.5, IQR: 64-71$). No one was colorblind.

\subsection{Procedure}
% \ming{TODO: describe the study procedure here: all the steps that we performed for the online user study.}
The eighteen older adult\rrv{s} \st{participants }participated in the study online from their homes using their own devices as they normally would do when using computers. 
Sixteen participants used their laptops, and two used their desktop computers.  
Based on the experience of the pilot studies, we moved some setup procedures into the \textit{pre-study} session so that participants could better focus on the tasks during the actual user study. 

\subsubsection{Pre-study Session}
We emailed participants the consent form before the study so that they would have sufficient time to read it and ask us questions. 
We sent the computer literacy questionnaire~\citep{boot2015computer} so that they could finish it before the study. 
\st{As}\rrv{Because} we conducted our study through Zoom, we scheduled with each participant to help them set up Zoom, if they had not already done so, and provided a short tutorial, such as how to share their screen.
The goal of the pre-study was to receive informed consent from participants and ensure that the technical setup was sound for the study.

\subsubsection{Online Think-Aloud Usability Testing}
%\ming{TODO: describe all the steps that we performed for the online think-aloud user study and the interview questions}
% The mixed-methods study included an online think-aloud usability testing with a set of interactive visualizations, interviews after each task, and questionnaires. 
We conducted the online think-aloud usability testing through Zoom and audio- and screen-recorded the study to understand how they comprehended the interactive visualizations\st{ through their think-aloud verbalizations}. 
Two research team members were presented in each study session.
One acted as the moderator to deliver the instructions and moderate the study. The other was the note taker who observed the session and noted down important observations.  

At the beginning of the study, the moderator explained the think-aloud protocol~\citep{Thinking18:online} and asked the participant to work on a task with an interactive visualization that would not be used in the formal study. The goal \st{was }of this process was to get each participant familiar with the think-aloud protocol. 
Next, the moderator presented the participant with the visualizations (Figure~\ref{fig:interactivevisualizations}) and the tasks (Table~\ref{tab:tasks-for-visualizations}). 
The order of the visualizations and the corresponding tasks were randomized. 
The moderator reminded the participants to keep talking if they felt into silence for a long period and monitored the time to remind participants to stop if the allocated time for each task was out. 
%After participants completed the task for each visualization or the allocated time was out, the moderator asked participants to rate how well they understood the visualization and how well the interaction was with two 7-point Likert-scale questions: \textit{Q1: I felt that I understood the visualization well}; \textit{Q2: I felt that I was able to interact with the visualization well}. 
% Moreover, the moderator conducted a semi-structured interview to further understand the challenges they encountered. 
% After participants completed all tasks, they were interviewed about their overall experiences with the visualizations.  
The order of the visualizations was randomized.
Each participant was compensated with \$25. 

% \subsection{Data Collection}
% \ming{TODO: describe different types of data that we have collected.}

\subsection{Data Analyses}
%\ming{TODO: describe how we analyze the data}
We audio- and screen-recorded the user study sessions. We transcribed the audios to acquire participants' think-aloud verbalizations.
%, which were the primary data of our analysis. 
%Next, we present how we analyzed each type of data.
%\subsubsection{Think-aloud verbalizations and screen recording Data}
We applied thematic analysis on think-aloud verbalizations. First, two coders independently coded the verbalizations to identify different comprehension thought processes. As think-aloud verbalizations may contain fragmented utterances~\citep{nielsen2002getting,charters2003use}, we referred to the screen recordings whenever needed to better understand participants' verbalizations during the entire analysis.  
% Specifically, the coders independently scrutinized whether participants' verbalizations fell into one of the three processes of the three-process visualization comprehension model (see Section~\ref{sec:VisualizationComprehension}). 
% If it did not fell into any of the processes, the coders would assign a new code to the verbalizations. 
Afterward, the two coders discussed their codes to gain a consensus on their codes. The third coder joined the discussion to resolve potential conflicts and consolidate the final codes. The coders \rrv{applied affinity diagramming to} group the codes into themes, which characterized different types of thought processes.
Finally, we compared the final themes of thought processes with the three-process visualization comprehension theory (see Section~\ref{sec:VisualizationComprehension}) to derive the commonalities and differences. 

We followed a similar approach to identify the challenges that participants encountered with the visualizations. We then organized the challenges based on the seven types of interactions~\citep{yi2007toward}. Findings will be presented in Sec~\ref{sec:challenges}.

% Findings will be presented in Sec~\ref{sec:thoughtprocesses}.

% We computed descriptive statistics and performed Friedman tests on the Likert-scale ratings~\citep{mackenzie2012human}. We also analyzed participants' interview feedback to understand participants' preferences and suggestions. Findings will be presented in Sec~\ref{sec:preferencessuggestions}.

% think-aloud and interview
% transcribe data
% two coders independently analyzed the data, identified themes
% two coders discussed and resolved conflicts. 

%\subsubsection{Interview Data}
% think-aloud and interview
% transcribe data
% two coders independently analyzed the data, identified themes
% two coders discussed and resolved conflicts. 

% \subsubsection{Quantitative Data}
% quantitative data:
% descriptive/statistical tests

%% file: 4-results.tex
\section{Findings}
%We present the results in three main themes to answer the RQs. 
% \ming{Two issues with quotes: 1) many have grammar issues; 2) many are incomplete and need additional contexts to understand. Thus, one key thing is to re-listen and check the quotes. Don't just rely on automatic transcription. I will add a mark at quotes that I think need to be revised.}

%\subsection{How do older adults comprehend interactive visualizations?}
\subsection{Comprehension \textit{Thought Processes} and the Associated \textit{Challenges} (RQ1)}
\label{sec:thoughtprocesses}
% \begin{table*}
%   \caption{The average number (percentage) of verbalized words in each type of thought processes among all participants for the five visualizations respectively and together.}
%   \label{tab:thoughtprocessesnum}
%   \begin{tabular}  {p{4.6cm}|p{1.6cm}|p{1.6cm}|p{1.6cm}|p{1.6cm}|p{1.6cm}|p{1.6cm}}
%     \toprule
%     Types of thought processes & A & B & C & D & E & All\\
%     \midrule
%     Encoding visual information  & 110 (18.71\%)  & 58 (18.13\%) & 114 (20.258\%) & 282 (19.18\%) & 110 (18.71\%) & 110 (18.71\%)\\
%     \hline
%     Relating visual features to concepts & 86 (14.63\%) & 31 (9.69\%) & 60 (10.68\%) & 177 (12.04\%) & 110 (18.71\%) & 110 (18.71\%)\\
%       \hline
%     Associating concepts with existing knowledge & 356 (60.54\%) & 219 (68.44\%) & 343 (61.03\%) & 918 (62.45\%) & 110 (18.71\%) & 110 (18.71\%)\\
%       \hline
%     Seeking help information &  20 (3.40\%)  & 9(2.81\%) & 21 (3.74\%) & 50 (3.40\%) & 110 (18.71\%) & 110 (18.71\%)\\
%       \hline
%     Selecting & 16 (2.72\%) & 3(0.94\%) & 24 (4.27\%) & 43 (2.93\%) & 110 (18.71\%)& 110 (18.71\%)\\
%           \hline
%     Error-correction &  20 (3.40\%)  & 9(2.81\%) & 21 (3.74\%) & 50 (3.40\%) & 110 (18.71\%) & 110 (18.71\%)\\
%           \hline
%     Task-induced processes &  20 (3.40\%)  & 9(2.81\%) & 21 (3.74\%) & 50 (3.40\%) & 110 (18.71\%) & 110 (18.71\%)\\
%   \bottomrule
  
% \end{tabular}
% \end{table*}

\rv{Our analyses uncovered four types of thought processes that older adults had when interacting with and comprehending the interactive visualizations: 1)\textit{Encoding visualization information}, 2)\textit{Relating visual features to concepts}, 3)\textit{Associating concepts with existing knowledge}, and 4)\textit{Recovering from errors}.} %, and 5) \textit{Comprehending the tasks}. 
\rv{The first three types of thought processes are consistent with the three-process comprehension theory discussed in Sec~\ref{sec:VisualizationComprehension}.
%Furthermore, we also uncovered two additional types of comprehension thought processes: .
% \ming{TODO: update the numbers in the table to show the number (percentage) of verbalized words in each thought process.}
% Table~\ref{tab:thoughtprocessesnum} shows the average number and percentage of verbalized words in each type of thought processes among all participants for the five visualizations. 
Furthermore, we also identified challenges associated with each thought process.} 

Next, we explain these four types of thought processes and the corresponding challenges in detail. We provide participants' think-aloud verbalizations as quotes and annotate each quote with \textit{``P\#, letter''} to indicate the corresponding participant ID and the corresponding visualization ID in Figure~\ref{fig:interactivevisualizations}. 

\subsubsection{Encoding visual information}

While comprehending visual characteristics of visualizations, participants encountered challenges with \textit{colors}, \textit{labels}, \textit{texts}, and \textit{layouts}.
Two issues were associated with colors: brightness and contrast. While it is common to use light-grey colors to indicate inactive or unchecked visual elements, our participants did not notice their existence in the first place, such as in Figure~\ref{fig:interactivevisualizations}.C: \textit{``My issue was trying to see something that wasn't there. There appears to be a very, very, very faint background graph''- P15, C.}
%such light-color designs caused difficulties for participants to notice their existence in the first place.  
% \begin{quote}
%     ``My issue was trying to see something that wasn't there. There appears to be a very, very, very faint background graph''- P15, C.  
%     % "Sometimes it was hard just hard to see the color variations. Spots you seem to get when I was looking It. Um... see the color faded. Can you see where the colors faded? It was hard to see what was going on." - P6, B, (24'41)
% \end{quote}

Participants also had difficulty perceiving differences between colors: \textit{``The difference between Europe and the Americas is not as clear as it is between Southeast Asia and Eastern Mediterranean.''- P15, B.}
When turning Figure~\ref{fig:interactivevisualizations}B into gray-scale, we noticed that Europe and Americas were indeed barely distinguishable.  %This was mainly caused by their visual acuity.  
% \begin{quote}
%     ``The difference between Europe and the Americas is not as clear as it is between Southeast Asia and Eastern Mediterranean.''- P15, Figure~\ref{fig:interactivevisualizations}B.
% \end{quote}

% We converted the Figure~\ref{fig:interactivevisualizations}B into the gray scale and noticed that the difference between Europe and Americas was indeed hard to discern. \ming{TODO: add a gray scale image of Figure~\ref{fig:interactivevisualizations}B}
%  - add GrayScale1.pdf
% o labels 
%     -visualizations lack labels (e.g., E)
%     -buttons also lack labels
%     -visual encoding (unfamiliar icon instead of text) (e.g., B); 

% \ming{TODO: the following quotes need to be revised to be clearer.
% -    the dotted line "in 3 days..." 
%             "I'm looking for countries over 10,000 deaths, So 10 is over here somewhere. Three days, two days, in five days, in one week. So I don't know how to..." - P14, E, (17'01)
%             "If you look at the thing there is a how many deaths in three days and five days and in a week. I'm not sure what those figures up in that corner are supposed to be used for." - P15, E.1, (40'11)
%             "I see other lines here on the far left, showing the rate of change and these upward sloping things, two times the number of deaths in one week or five days. This looks like a chart that might be useful for professional in in public health. Not here for the average person" - P19, E.1, (35'37)
%             }
%     \ming{ TODO: quotes; and crop out the 3 lines areas;)}

The key issue with labels was the lack of them. This happened to legends and axis labels.
For example, Figure~\ref{fig:interactivevisualizations}D did not have a legend to explain what COVID data the circle size represents for: \textit{"I just didn't know what the circle was referring to, was it a place?"-P7, D}; \textit{"The diameter of the circle, what they represent is not clear." -P10, D.}
Second, the axes of some visualizations lacked necessary labels. Figure~\ref{fig:interactivevisualizations}E.1 did not have labels for both its axes. This caused the misinterpretation of data. \textit{"So the hundred [a number on the horizontal axis] must be days [it actually referred to the number of deaths]. But why the hell they didn't say that." - P13, E.} 

Two issues were associated with texts in the visual. First, some font sizes were perceived too small, such as texts in Figure~\ref{fig:interactivevisualizations}A.1. Second, some texts lacked sufficient brightness or clear borders. \textit{ "I think they should make the printing a little brighter, so I can read it." - P6, C.}

% o position [visual clutter/overlap (e.g., D,A, bubble map)]
%     -Lines/Bubbles overlapping (e.g., E (Lines of Germany and Canada), A (Bubble's borders are overlapping in map))
%         -some realized the two lines were overlapped but some did not realize it
%             -magnifier 
%     -information should always be shown instead of just being shown when hover over
%         -hover over is not always sufficient
%             -WHO: 
%         -lack of the select interaction: anchor
One challenge with layouts was overlaps between visual elements. %Overlaps between visual elements was the key challenge associated with layouts. 
%Participants were confused about what each of these overlapped visual elements represented for. 
For example, Figure~\ref{fig:interactivevisualizations}D.2 had many overlapped circles. 
Some participants even thought the overlapped visual elements were a single element. Many countries' lines were \rrv{too}\st{very} close to \rrv{each other} or partially overlapped with each other in Figure~\ref{fig:interactivevisualizations}E, such as the ones for Germany and Canada. Some participants did not even realize there were two lines. 
Moreover, the sheer amount of visual elements could also be overwhelming. \textit{"You ask me to find one line, but there are too many lines" - P1, C.}

\subsubsection{Relating visual features to concepts.} 
\label{sec:relatingvisualfeatures}
This type of thought process included cognitive operations creating a mapping from visual features to conceptual relations represented by those features.
In general, participants were able to relate the continuous changes in visual features to the trends of underlying data. For example, participants intuitively associated different shades of blue in Figure~\ref{fig:interactivevisualizations}A and different sizes of the bubbles in Figure~\ref{fig:interactivevisualizations}D with the severity of COVID conditions. 
%Despite it was easy to interpret trends represented by the continuous changes in visual features, 
However, participants often had difficulty understanding mappings between visual features and the underlying COVID information they represented.
Think-aloud verbalizations revealed three reasons for such difficulties.

% \begin{quote}
%           - "I just didn't know what the circle was referring to, was it a place?"-P7, D (1'00'10)
%       - "The diameter of the circle, what they represent is not very clear." -P10, D (56'27)
% \end{quote}

% For example, it was unclear whether the bubble size in Figure~\ref{fig:interactivevisualizations}(D) represented the total death or the total cases. 
%  \begin{wrapfigure}{R}{0.25\columnwidth}
%   \centering
%     \includegraphics[width=0.25\columnwidth]{Figures/value-log.PNG}
% \end{wrapfigure}
 \begin{wrapfigure}{r}{0.25\columnwidth}
  \centering
    \includegraphics[width=0.25\columnwidth]{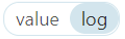}
% \Description{A toggle button of Figure C.1 labeled "value" and "log" to switch the graph type}
\end{wrapfigure}
\textbf{Ambiguous Language}: The language used in some visualizations was ambiguous. Three factors contributed to ambiguity. 
The first factor was \textit{use of technical terms}. For example, participants were confused about the meaning of the term \textit{``log''} in the inline figure on the right (i.e., the top-left corner of Figure~\ref{fig:interactivevisualizations}C.1). 
Similarly, participants also had hard time making sense of the terms \textit{``trends''} and \textit{``count''} in Figure~\ref{fig:interactivevisualizations}E, which were used for the ``log'' and ``linear'' scales. 

Another factor was the \textit{use of abbreviations}. 
For example, the labels on the horizontal axis used the letter \textit{``m''}. Although it was intended to mean ``million,'' some participants misunderstood it as ``month.'' 
Such misunderstanding was reasonable because it was not uncommon to show time on the horizontal axis. 
%  \begin{wrapfigure}{R}{0.16\columnwidth}
%   \centering
%     \includegraphics[width=0.16\columnwidth]{Figures/trends-counts.PNG}
% \end{wrapfigure}

The last factor was \textit{use of incomplete labels}.
Short words or phrases are often used in visualization labels to maintain a ``clean'' looking as complete descriptions might make visualization cluttered. 
However, such incomplete labels caused confusions. 
For example, it was unclear to some participants what the word \textit{``total''} in Figure~\ref{fig:interactivevisualizations}A referred to because it could mean the ``total'' confirmed cases, ``total'' new cases, or even ``total'' global cases.  
Similarly, it was also unclear whether the word ``cases'' in Figure~\ref{fig:interactivevisualizations}B referred to the cumulative or daily cases.
% \molly{Also, one participant mentioned she expected "Total" is for the total of global case, which cause further confusion on the options inside of dropped down menu. She perceived there should be data from different countries. "This is for global cases and it will shows details, now there is new cases, confirmed cases and deaths. If I click it, shows the cases from different countries and regions, like US's cases" - P17, A.1}

\textbf{Mental model mismatches:} \rrv{Confusions were also caused by }two types of mismatches between participants' mental models and the information organization in the visualizations\st{ also caused confusion}. 
One mismatch was related to the \textit{grouping of options in the visualization}. For example, Figure~\ref{fig:interactivevisualizations}A organized the ``total per 1 million population'' option into the drop-down list with the default option as ``total.'' 
However, many participants did not expect that the ``total per 1 million population'' option belonged to the same category as the ``total'' option.

Another mismatch was related to \textit{the mappings between graphical components}. 
For example, the slider-bar at the bottom-left corner of Figure~\ref{fig:interactivevisualizations}E allowed users to change the number of \textit{deaths} to change the number of \textit{days} displayed on the x-axis of the visualization. 
However, such mapping was counter-intuitive, and many participants were confused about whether they actually changed the number of deaths or days.   

%  \begin{wrapfigure}{R}{0.25\columnwidth}
%   \centering
%     \includegraphics[width=0.25\columnwidth]{Figures/affordance-icons.png}
% \end{wrapfigure}
 \begin{wrapfigure}{r}{0.25\columnwidth}
  \centering
    \includegraphics[width=0.25\columnwidth]{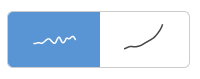}
% \Description{A toggle button of Figure1B labeled one curvy line and one steady line to switch between daily change and cumulative}
\end{wrapfigure}
\textbf{Affordance mismatches:} Confusions also occurred when participants' perceived affordance of visual elements did not match with what they represented. 
For example, a ``smooth'' and a ``wigging'' line icons in the inline figure (part of Figure~\ref{fig:interactivevisualizations}B.4) were used to represent the ``cumulative'' and ``daily change'' cases. Unfortunately, no participants perceived the two line icons to be related to these two terms.

% \begin{quote}
%          - "This is confusing for me [she points to the two line icons]. Because it[line icon] is all the waviness, not the bar graph. This is bar graph[she points to the B.2] and this is not the bar graph[she points to the line icon], this[line icon] is the computer graph." - P5, B (41:03)
%          - "I would guess from the graphic that it meant added together but I mean why not just put that word in there(he points to the line icons). Because that word is there(he points to the Daily and Weekly icons). I mean, that was other words are there." - P14, B (51:58) 
% \end{quote}

%  \begin{wrapfigure}{R}{0.25\columnwidth}
%   \centering
%     \includegraphics[width=0.25\columnwidth]{Figures/EyeDatatable.png}
% \end{wrapfigure}

% The visual appearance of the dropdown list that contained the ``Total'' option in the floating tool bar in Figure~\ref{fig:interactivevisualizations} A.1 caused confusions.
 \begin{wrapfigure}{r}{0.258\columnwidth}
  \centering
    \includegraphics[width=0.258\columnwidth]{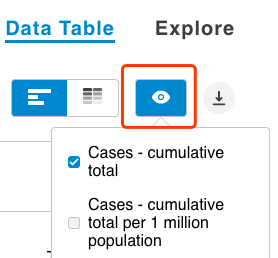}
    % \Description{A eye button inside of Data Table tab in Figure 1A shows several attributes related to COVID-19, such as cases of cumulative total, cases of cumulative total per 1 million population, and so on. The color of active status is blue, and the white is for inactive}
\end{wrapfigure}

Another common perceived affordance issue was the misuse of buttons and drop-down lists. 
For example, although the visual element with an ``eye'' icon in the ``data table'' tab in the inline figure here (part of Figure~\ref{fig:interactivevisualizations}A) functioned as a drop-down list, it visually resembled a button. 
Consequently, many participants did not expect it to show a list of options once clicked.

Moreover, the affordance of an ``eye'' often suggested revealing some hidden content, such as a password, which was also unrelated to a drop-down list. \textit{``The eye on some of the software I use ... allows me to see my password.'' - P11, A}. 
% \begin{quote}
%     " I had to really kind of fart around to figure out that that little I was the icon that I wanted in order to be able to get additional information which wasn't showing up on the chart." - P11, A, (15'58)
    
%     "I don't think that little icon of the eye is necessarily very intuitive to the fact that that is looking at it when the eye on some of the software that I use, it is to show information that like your password if you wanted to see your password. The little eye allows you to see your password. I do know that there's information that I can't see." - P11, A, (17'54)
    
% \end{quote}

%  \begin{wrapfigure}{R}{0.25\columnwidth}
%   \centering
%     \includegraphics[width=0.25\columnwidth]{Figures/slider-bar-affordance.PNG}
% \end{wrapfigure}

% Another common affordance issue was to discern the active tab of a toggle button. 
%  \begin{wrapfigure}{r}{0.15\columnwidth}
%   \centering
%     \includegraphics[width=0.15\columnwidth]{Figures/slider-bar-affordance.PNG}
% \end{wrapfigure}
Although it was common to use different colors to distinguish the active and inactive tabs in a toggle button, \rrv{participants were confused about} \st{it was confusing for participants to recognize} which color represented the active tab. For example, the blue ``Daily'' tab of the toggle button in Figure~\ref{fig:interactivevisualizations}B.4 was supposed to be the activated one. However, many participants felt that the white ``weekly'' tab was activated because it seemed to be protruding up.   
% As there could only be one active option, it was easier to infer the active tab when there were three or more options in the slider-bar. 
While it might be possible to infer the active tab by clicking each option and observe the change in the visualization, such a trial-and-error approach was not widely used by participants.  

Furthermore, affordance issues also occurred with the \textit{stacked bar} chart. 
Many participants perceived all bars in the stacked bar chart in Figure~\ref{fig:interactivevisualizations}B.2 starting from the same horizontal axis, and shorter bars simply overlapped on the top of tall bars. 
%Participants felt the taller bars were overlapped by the short bars.
\textit{"I thought the Americas was starting on the same axis as all the others. It just shows that how the Americas went higher. " - P4, B}.  
Consequently, as Americas was always shown as the top portion of the stacked bar chart, many participants thought that Americas' cases were always higher than other continents. 
As a result, they were confused when noticing that the number for Europe was higher than that of Americas in some parts of the chart. 
% One possible reason for the mismatch between the perceived and intended affordance of the stacked bar chart might be that the horizontal bars showing the numbers of different continents in Figure~\ref{fig:interactivevisualizations}B.1 were all left-aligned. This might have given the impression that the bars on the right side (Figure~\ref{fig:interactivevisualizations}B.2) must also be aligned.  
% \begin{quote}
%   \ming{any quote to illustrate the above point?}
%   Updated:  "I've put into short term memory that yellow is America's and green is Europe, so I don't, I'm not reading that I'm just looking at the numbers, changing But when I'm not highlighted when I'm looking like that, then I'm looking at the chart, obviously." P14, B, 48'45
% \end{quote}

\subsubsection{Associating concepts with existing knowledge}
\label{sec:associatingconceptswithexistingknowledge}

Participants tended to use their familiar tools, personal knowledge, and prior experiences with similar products or to follow general design principles when comprehending the visualizations. Confusion arose when visualization designs conflicted with their prior knowledge and experiences.

\textbf{Familiar tools}: General tools, such as Google and common commands in browsers, were used in their comprehension processes. 
For example, to find out the confirmed cases ``per 1 million population'' in France in Figure~\ref{fig:interactivevisualizations}A, many participants first found the total cases from the visualization, then searched in Google to find out the population of France, and finally calculated the number themselves. 
Although the map visualizations had + and - icons for zooming, some participants still chose to zoom in the entire website. 
What's more, we also observed that some participants used \rrv{common} hotkeys, such as Ctrl+F, to search for information in the visualizations.      

% \textbf{Personal knowledge}: Personal knowledge also moderated their comprehension processes. For example, some participants did not need to spend much effort in zooming in the world map and panning it to find France as they already knew the location of France. \molly{While asked which countries have over 10K cases, one participant came up with some potential countries first, and then check in the graph to see if the existing knowledge can match with the graph provided}

\textbf{Prior experiences with similar software}: Participants' prior experiences with similar software also affected their expectations of the visualizations. 
When clicking on the state name in Figure~\ref{fig:interactivevisualizations}D.2, they expected that would lead to more details of the state based on their prior experience with a similar website. \textit{"If I click [states on the map], do they go anywhere? No... So the New York Times site has, like, if you click Tennessee, it will go to Tennessee." - P14, D}.
% \begin{quote}
%     - "If I click them. Do they go anywhere. No. So the New York Times site has, like, if you click Tennessee, it will go to Tennessee." P14, C, (07'15) 
% -Updated below:
%     "I don't know if I click on one of these(counties on the map), does it give me more information? No... If we scroll down(the page), states listed. If I click them, do they go anywhere? No... So the New York Times site has, like, if you click Tennessee, it will go to Tennessee." - P14, C, (07'15)
%     \ming{TODO: update the quote to make it complete and/or remove grammar issues.}
% \end{quote}
Moreover, when hovering over a line in Figure~\ref{fig:interactivevisualizations}C, participants expected to see more information popping up. \textit{"Often in this kind of tool when you hover over the line. It tells you what you're looking at. But this one does not." - P14, C}. 
% \begin{quote}
% -Updated below:
%         "Often in this kind of tool when you hover over the line. It tells you what you're looking at. But this one does not." - P14, C (58'01)
%             \ming{TODO: update the quote to make it complete and/or remove grammar issues.}
% \end{quote}
Similarly, when zooming in the world map in Figure~\ref{fig:interactivevisualizations}A, participants expected more information would pop up immediately, just \st{as }like Google Maps. However, the world map did not reveal any further information after the first click on the ``+'' icon in Figure~\ref{fig:interactivevisualizations}A.5.  

%  \begin{wrapfigure}{R}{0.4\columnwidth}
%   \centering
%     \includegraphics[width=0.4\columnwidth]{Figures/Consistency_1.pdf}
% \end{wrapfigure}
 \begin{wrapfigure}{R}{0.5\columnwidth}
  \centering
    \includegraphics[width=0.5\columnwidth]{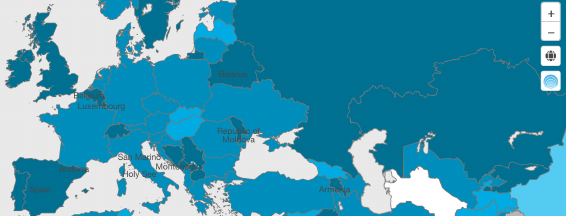}
        % \Description{A part of world map in Figure 1A colored by different level of blue and some countries' name show up like Spain, San Mario but some countries doesn't show}
\end{wrapfigure}

\textbf{General design principles}:
General design principles guided participants' expectations and comprehension processes. 
One such principle was \textit{consistency}. 
For example, when zooming in the map shown on the right (a part of Figure~\ref{fig:interactivevisualizations}A), not all countries' names appeared at the same time. This design was inconsistent with their experiences of using common digital maps, such as Google Maps. As a result, it confused participants as they would expect the country name to appear at the same time.

Another consistency issue was caused by a mismatch between the perceived meaning of \textit{an icon label} and \textit{the visualization it represented}. For example, two line icons were used in the toggle button in Figure~\ref{fig:interactivevisualizations}B.4 to represent the bar charts in Figure~\ref{fig:interactivevisualizations}B.2. \textit{"This is confusing for me [pointing to the two line icons] because they are lines, not the bar graph [pointing to the bar graph in Figure~\ref{fig:interactivevisualizations}B.2]. " - P5, B}.
 \begin{wrapfigure}{R}{0.24\columnwidth}
  \centering
    \includegraphics[width=0.24\columnwidth]{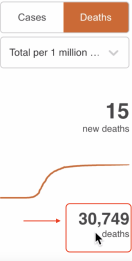}
    % \Description{Dashboard of Figure1A has a toggle button colored by dark orange to switch cases and deaths, a selection box labeled Total per 1 million and two numbers below the chart which is the number of new deaths and deaths}
\end{wrapfigure} 
Furthermore, the violation of consistency guidelines also happened when icons and textual labels were mixed together in the same group of visual elements. For example, while one toggle button in Figure~\ref{fig:interactivevisualizations}B.4 used icons as labels, the other two toggle buttons used texts as labels. Participants felt puzzled about the meanings of the icons and also the inconsistent design. \textit{"Why not just put words in there [pointing to two line icons]? 
Because words are used there [pointing to the Daily and Weekly icons]." - P14, B}.

Another design principle violated was \textit{Gestalt principles}. Gestalt principles are principles of human perception that describe how humans group similar elements and recognize patterns~\citep{koffka2013principles}.
One of these principles was the law of proximity. 
When two visual items were separated, they would not be perceived as connected. For example, since the tool bar on the left (A.1) was distant from the world map in Figure~\ref{fig:interactivevisualizations}A, many participants did not realize that the tool bar was connected with the map.
\textit{"Maybe it [the tool bar] can be placed around here, closer to the map... if it is closer, then it becomes more obvious [that it is part of the map]. " - P10, A}.
Consequently, when selecting a different option from the dropdown list in the tool bar, many did not notice that the colors in the map changed accordingly.
Similarly, the search bar at the top (A.2) was so far away from the map in Figure~\ref{fig:interactivevisualizations}A that many participants did not realize that it was connected to the map.

On the other hand, when two unrelated visual elements were placed too close together, they would be mistakenly perceived as related. 
For example, the drop-down menu (i.e., ``Total per 1 million...'') shown on the bottom right here (a part of Figure~\ref{fig:interactivevisualizations}A.1) was closer to the numbers below (i.e., 30,749) than the two buttons above (i.e., the ``Cases'' and ``Deaths'' buttons). As a result, many participants mistakenly thought that the number meant the number of deaths per 1 million population instead of the number of total deaths. \textit{"Does this [Pointing to the number] mean there are 30,749 deaths per million? Or is that the total of 30,749 deaths in France?" - P16, A}. 

\subsubsection{Recovering from Errors}
%\ming{TODO: thought processes related to recover from an error (e.g., backtracking to a previous step when encountering a problem)}
%When participants encountered problems when comprehending visualizations, they often attempted to correct and recover from issues. 
% One common cause for errors was that visualizations did not response immediately to users' actions. This could be caused by internet speed or their computer configuration. Participants often interpreted delayed reactions to their input, such as click, as ``not working''. For example, after clicking on Nevada in Figure~\ref{fig:interactivevisualizations} D, P19 did not see the map zooming in and thus moved away from the map.  
% Another cause for errors was glitches in visualizations. After clicking one country name in Figure~\ref{fig:interactivevisualizations}E, all functions suddenly disappeared. P6 tried to scroll the page to find the disappeared visualization.  
Some of the participants' thought processes were related to figuring out and recovering from errors. 
%\textit{"The problem is that I clicked on France, and now it's zoomed in. [attempting to recover by clicking on the map] But I can't get it to zoom back out" -P15,B}. 
One error-recovery approach was \textit{trial-and-error}. For example, when trying to figure out why Germany was not added in E.4 for the task on Figure~\ref{fig:interactivevisualizations}E, P2 attempted different options in E.3 and verbalized: \textit{``...Trends, Counts...[clicking the Counts button]... Should be a different graph? Let's see what happens'' - P2, E}.

Another error-recovery approach was to adopt their familiar non-digital workarounds. For example, when P11 could not figure out how to add multiple countries in Figure~\ref{fig:interactivevisualizations}E at the same time, he eventually decided to add one country at a time and wrote down the country name and the case number shown in the figure on a piece of paper. Later, he compared these written numbers to complete the task.

\subsection{Challenges with Interaction Techniques (RQ2)}
\label{sec:challenges}

% \subsubsection{Internet connectivity and devices}:
% \ming{TODO: add examples that were caused by connection and device issues}

%     - "I'm going to try click on Nevada. And see if anything happens. It's not working and not getting any actions." - P19, D, (2'12)
%     - "So I see down here select up to three countries to compare[E.4] and it looks like I can X out these (Mainland China and United States)...I don't know Why it's not working... Here[clicking word "countries" in "Select up to 3 countries to compare..."] and see what happens. I'm gonna wait. Aha, it's just my computer is a little slow on my laptop." - P19, E, (16'16)
% \yuni{Done}

% - Device/Internet connection problem affect the understanding (lagging)
%     -problem?
%         -example of an issue that was caused by poor internet connectivity
%         -feedback to remind them that the visualization was still working 
%         - "when I clicked on the state[Nevada] itself, cause I didn't see the icon at the top[loading icon on the tab] that was loading. I thought nothing was happening." - P19, (5'41)
% \yuni{Done: Actually there was no loading icon shows either on the tab or on the map after the participant clicks on Nevada state.}

% think-aloud data + interview data + observation + videos

% quantitative data table 
% how many participants encountered problems with each interaction technique?

% seven interaction techniques 
\rv{In addition to understanding participants' comprehension thought processes, we analyzed the challenges they encountered with the seven interaction techniques (see Section~\ref{InteractionTypes}). We found that they encountered problems primarily with the following five interaction techniques: Abstract/Elaborate (e.g., hover-over tooltips and zooming), Select (e.g., pinning a visual element and making it always visible), Filter, Encode (e.g., visual encoding), and Reconfigure (e.g., changing spatial arrangement of underlying data).}

% The occurrences of the issues associated with these eight interaction techniques is as follows: Abstract/Elaborate (X), Filter (X),   

\subsubsection{Abstract/Elaborate}

 \begin{wrapfigure}{r}{0.55\columnwidth}
  \centering
    \includegraphics[width=0.55\columnwidth]{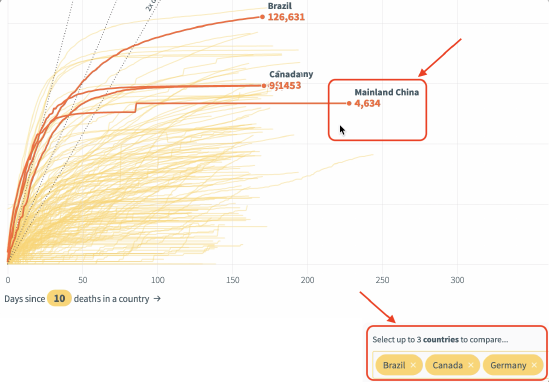}
        % \Description{Figure1E Line chart has several highlighted lines colored by dark orange, which are Brazil, Canada, and Mainland China. The light orange represented other countries worldwide. One Selection box is in the right side below the main line chart, which labeled Select up three countries to compare and Brazil, Canada, and Germany have been selected}
\end{wrapfigure}
Hover-over tooltips and zooming were two examples of Abstract/Elaborate. 
A common issue with hover-over tooltips was precise pointing and selection. 
Selecting an item from many closely arranged visual elements was challenging. 
For example, stacked bars in Figure~\ref{fig:interactivevisualizations}B.2 were so close together that a slight motion slip resulted in a different bar than the intended one.
Hovering over thin lines, such as the ones in Figure~\ref{fig:interactivevisualizations}C, to reveal tooltip was an issue for participants as this required them to move their mouse with fine motor control.

Accidentally triggered hover-over tooltips caused confusion. For example, as is shown in the inline figure here (a part of Figure~\ref{fig:interactivevisualizations}E), after selecting the three required countries (i.e., Brazil, Canada, and France) as indicated in the bottom right corner, P2 was surprised to find that another country's data was also shown up and became confused: \textit{"Why am I getting China?"}. 
%\molly {We have the same code in Select section as well} 
This was because he did not realize that his cursor was still on the chart and the cursor triggered a pop-up tooltip at that place, which displayed the data for China.

Participants anticipated zooming in on a map would reveal more information similar to how Google Map works. However, zooming in on a country in Figure~\ref{fig:interactivevisualizations}A did not reveal more details.
%some were confused when they realized no details were shown when they zoomed in the map in Figure~\ref{fig:interactivevisualizations}A. 
Another issue with zooming was that the ``+'' and ``-'' buttons were so far away from the map in Figure~\ref{fig:interactivevisualizations}A that they were not perceived to be related to the map and thus were rarely used. This was the violation of Gestalt principles (see Sec~\ref{sec:associatingconceptswithexistingknowledge}). 

% % - Zoom-in
%   o Zoom-in and details
%   - Looking for information pop up after zoom-in - P4，A （2:09:00 & 2:13:58） 
  
%   - " Because I thought that the map would allow me to enlarge it enough to find France, to click France. Yeah, that would give me the France information." - P4, A, 2'13"58
%     - "So if I zoom in[the map], would it give me more information? No."- P13, A, 32'22
    
%     - Assume the zoom-in the area of France would show more details. - P4, A (2:21:13)
%     - Fail to relate the scroll bar with the zoom-in function(E)
% ----------------------------------------------------------------    

\subsubsection{Select}
While Abstract/Elaborate (e.g., hover-over tooltip) can reveal additional information, Select techniques allow for marking a visual element and pinning associated information to facilitate comparison. The task for Figure~\ref{fig:interactivevisualizations}B required participants to compare the numbers between Americas and Europe. 
%Although the hover-over tooltip could reveal the number for Americas or Europe respectively, the number would disappear once the cursor was moved away.  
Participants hoped to click on the bars representing Americas and Europe to select/pin them there so that they could see numbers for both at the same time. Unfortunately, Figure~\ref{fig:interactivevisualizations}B did not support Select techniques.  

Although Figure~\ref{fig:interactivevisualizations}C and E supported Select techniques by allowing for clicking a country option in C.1 to \rrv{activate}\st{active} the corresponding line in C.2 or adding a country in E.4 to highlight the corresponding line in E.1, such Select techniques were realized through separated UI elements (i.e., C.1 and E.4 respectively) outside of the main visualization areas. However, participants tended to directly click on the lines that they want to activate. In other words, Select techniques could be improved by supporting \textit{direct manipulation} on the visualization instead of relying on a UI element physically separated from it.  
% -    	Select: While hover over need more mobility, select is necessary to highlight

    % -design trade-off between Select and Hover-over
    %      -Elaborate/Abstract (hover-over) might distract Select
    %          - "Why am I getting China?" - P19, E \molly {Same code with Hover over!!! 
    %          Suggested to replace with "But the interacting with it, I'm having a lot of trouble. As I move the cursor, keeps bringing up different countries. So that's very confusing, because I'm trying to read it. And then if I hit the cursor, it keeps bringing up different countries. -P7,E,1, (19'07)
             
    %          According to the feedback, this participants tried two interaction techniques during the process, however, he found that by hitting the cursor (Select), the result is same as moving the cursor. Apparently, the design didn't balance well between the select and hover over.}

    %          - "What are showing here give me a mainland China, especially in Germany, we know the mainland China, South Korea, it's not showing Germany." - P20, E
    
    %      -Select is necessary and better than hover-over (lack of the select interaction: anchor) 
    %         -participants wanted the information to be there and did not want it to disappear (hover-over)
    %             -Figure1.B.2 

\subsubsection{Filter} 
While participants were familiar with standard drop-down lists, many did not realize that they could filter options in the country drop-down list in Figure~\ref{fig:interactivevisualizations}E.4. As a result, they scrolled through the long list of country names to find the target. One common issue with scrolling a long list was that the list would disappear if their cursor was accidentally moved out of the list area. As a result, they were confused and had to start over again. 
Another issue was caused by the misalignment between the visual indicator of the selected item in the list and their mouse cursor, which might be caused by poor internet speed or computer configuration. 
%For example, the highlighted item in the list and the cursor was out of sync for some participants in Figure~\ref{fig:interactivevisualizations}E.4, which caused them to select the wrong item. 

Some filters lacked real-time feedback. For example, while dragging the scroll bar in Figure~\ref{fig:interactivevisualizations}E.2, participants were confused that nothing happened in the visualization. This was because the visualization was only updated when the mouse was released. 

Furthermore, similar to their expectation for Select techniques, participants also expected to use Filter techniques by \textit{direct manipulation}. For example, although the scrollbar in Figure~\ref{fig:interactivevisualizations}E.2 allowed for filtering lines in the visualization, few participants used it. Instead, many participants directly clicked on a number label in Figure~\ref{fig:interactivevisualizations}E.1 and anticipated it to filter lines in the visualization based on the number.

\subsubsection{Encode}
Participants found the visual encoding of some visualizations ambiguous and hard-to-understand and consequently attempted to interact with the visualization in the hope of changing the visual encoding to better understand the data. \textit{``I see a bubble in the place where Germany, Italy, and Switzerland joined, so is the statistics based on Italy, Germany, or Switzerland? It’s hard to tell... [click A.1 to switch from bubble map to choropleth map]" - P8, A}. 

% - Encode different types of graph: 
% "And use the bubble to make a map, and then the color blocks. I think it's too cumbersome. I don't need so many things. It doesn't seem to be very interesting. What is the difference between using the bubble-like things and color blocks to express the essence? I only think that bubbles seem to have a kind of diffusion, but you can’t count them with this bubble method. For example, there is a bubble in the place where Germany, Italy and Switzerland joined, so is your statistics based on Italy, Germany, or Switzerland? It’s hard to tell, right?" - P8, A, (1'18"40) \yuni{Translated this participant's quote into English}

% Thinking process:
% "So I'm going to leave this bubble thing because I don't like it. I'm going to go back to the Choropleth? I can't quite see, it is too small. As you can see I wear glasses. And I've still got deaths selected. And so it's now different? Okay, I got it. The different shading of the countries represents the deaths but what this map isn't telling me is what the different shades of brown represents. I don't know if a dark brown is worse or better." - P11, A, (23'13)

% - D switching between types of coding schemes was confusing (bubble to color) 
%     - P8,D(TODO) \yuni{This participant doesn't like the bubble and color because it is too cumbersome.}

% - Encode different types of graph: 

%     - the bubble map along with the region map, too complicate 1'18'40 But the bubble is confusing especially in the border of two different countries. Can't distinguish the country particularly - P8

\subsubsection{Reconfigure}
Reconfiguration allowed participants to change the spatial arrangement of underlying data (e.g., sorting and rearranging columns or changing x- and y-axes). Figure~\ref{fig:interactivevisualizations}B and E allowed participants to use slider bars (B.4, E.3, and E.4) to reconfigure data shown in different axes. 
However, participants found that it was not always clear which option in a slider bar was activated. One potential reason was mismatched affordance, which has been explained in Sec~\ref{sec:relatingvisualfeatures}.

%% file: 5-discussion.tex
\section{Discussion}
We understand how older adults comprehend COVID-19 interactive visualizations through the lens of their think-aloud verbalizations. In this section, we discuss these thought processes in the light of the literature and the challenges \st{that older adults }that older adults encountered when using \st{with }these interactive visualizations. 

\subsection{Visualization Comprehension Thought Processes and Challenges}
\subsubsection{Thought Processes of Comprehending Interactive Visualizations} 

Our analyses uncovered four main types of thought processes that explained how older adults comprehended the COVID-19 interactive visualizations: \textit{encoding visual information}, \textit{relating visual features to concepts}, \textit{associating concepts with existing knowledge}, and \textit{error-recovery}
%, and \textit{Task-induced} processes. 
Older adults experienced an integrated and iterative cycle of these thought processes.

Previous research studied how \textit{young adults}, mostly college students, comprehend \textit{static visualizations} and identified the first three types of thought processes: encoding visual information, relating visual features to concepts, and associating concepts with existing knowledge~\citep{shah2002review,carpenter1998model,barnard2013learning}. 
Our research extends this line of research by studying how \textit{older adults} interact with and comprehend \textit{interactive visualizations} that are closely related to their lives. We found the same three types of thought processes. \st{What's more}\rrv{Moreover}, our research further revealed one additional type of thought processes: \textit{error-recovery}.
%, and \textit{Task-induced} processes. 
These four types of thought processes provide a lens to understand how older adults comprehend information from interactive visualizations at a micro-level. 

\subsubsection{Challenges in Comprehending and Interacting With the Visualizations}
Our study revealed challenges older adults encountered with each type of thought process (see Sec.~\ref{sec:thoughtprocesses}). These challenges were caused by inappropriate visual designs (e.g., colors, texts, layouts), insufficient information (e.g., lack of labels, missing legend), confusing information (e.g., unfamiliar labels, ambiguous language), mismatches in mental models as well as in affordance, conflicts with prior knowledge and experiences, and violations of design principles. 
\rv{While prior research focused mostly on static visualizations, our work focused on interactive visualizations and extended our understanding of how older adults perceive and use interactive visualizations.}
\rv{For example, prior work found that static stacked bar charts takes longer for older adults to comprehend than bar charts~\citep{le2014elementary}. In contrast, our study found that dynamical rearrangement of stacked bars in a bar chart after a user interaction can affect older adults' understanding of the chart.}

%\rv{Our work primarily focuses on the user scenario of dealing with interactive data visualizations compared with previous literature related to the visual challenges encountered by aging population. For example, we discovered the affordance issue of arranging portions of the stacked bars affects the accuracy of understanding, which supports previous findings as comprehending stacked bar charts take longer time than bar charts ~\citep{le2014elementary}.}

Furthermore, our study found that older adults encountered challenges with five of the seven types of interaction techniques (see Sec~\ref{sec:challenges}). 
Specifically, they encountered more challenges with Abstract/Elaborate, Select, and Filter techniques than with Encode and Reconfigure techniques. 
These challenges were primarily caused by inadequate responsiveness of interactive visualizations (e.g., delay between user actions and visual feedback), demand of precise pointing (e.g., being able to move cursor in a fine-granular way), and unfamiliar or confusing visual elements (e.g., the filter within a text box). \rv{Such challenges might be exacerbated with aged-related perception and motor declines, such as presbyopia and finger coordination~\citep{Agerelatedchanges}, and thus are more likely to appear among older adults than young adults. However, more controlled comparative studies are needed to validate this conjecture.}
%\rv{One possible reason might be that all these interaction techniques require fine-grained pointing and selection operations with a computer mouse, which can be challenging for older adults as prior research suggested that older adults are more likely to have presbyopia and less control of hand presbyopia ~\citep{Agerelatedchanges}.} 

Our analyses also revealed ways to improve the user experience of interaction techniques commonly used in interactive visualizations: balance complementary interaction techniques (e.g., Abstract/Elaborate and Select), consider using direct manipulation that allows for moving visualization directly on it instead of relying on controls physically distant from the visualization, match the perceived affordance of a UI element with the function that it supports, and allow for alternative ways of exploring visualizations to increase the chance of success. 

\subsection{Design Guidelines}
Based on our findings, we derive five design guidelines (DGs) for creating user-friendly interactive visualizations for older adults. We further discuss how these DGs corroborate with and extend the literature as well as how they might be implemented.

\textbf{DG1: Increase the legibility of visual designs}. This guideline entails three sub-guidelines. First, the color scheme used in the visualization should have sufficient contrast. Second, texts should have sufficient size and contrast. Low contrast color schemes and small font sizes can be hard for aging eyes. \rv{These two sub-guidelines are consistent with the general guidelines for creating user friendly instructions for older adults~\citep{fan2018guidelines}. Previous research suggests that the font size should be at least 12-point to accommodate aging vision~\citep{becker2004study,chadwick2002web,chisnell2004designing,bernard2001determining}. However, these prior works focused on static charts on paper or screen. In contrast, our work shows that these guidelines are also applicable to interactive visualizations.}

Lastly, visual elements representing important information should be spaced out properly. The densely-distributed lines in Figure~\ref{fig:interactivevisualizations}E and bars in Figure~\ref{fig:interactivevisualizations}B represent many data visualizations commonly found online. Although such visualizations might not cause severe issues for young adults who have fine motor skills and good eyesight to explore the meaning of each line or bar by moving their cursor from one to another, our study shows that older adults struggled to do so and often overshot the intended line or bar. Thus, visual elements (e.g., lines or bars) representing important information should be spaced out generously so that it does not require fine motor skills and good eyesight to navigate them. \rv{While prior work suggests that this guideline is necessary for designing user interfaces on a small mobile phone for older adults~\citep{de2014design}, our work extends the prior work and shows that such guideline is also necessary for designing interactive visualizations for older adults.}

\textbf{DG2: Facilitate the understanding of the affordance of visual elements by adding textual labels and using plain language}. First, important interactive visual elements, such as buttons, should be labeled with text even when icons are used. \rv{While prior work shows the necessity of this guideline for mobile user interface design, our study suggests that icons alone were not always interpreted correctly by older adults with the intended affordance and thus text labels should be added to assist the understanding of interactive visualizations~\citep{de2014design}. \rrv{Moreover, our study shows that short labels can be misinterpreted and cause confusion. Thus, visualization designers should balance between keeping texts succinct to maintain visual aesthetics and informative to make it self-explanatory.}} 

Second, plain language should be used to explain technical terms or abbreviations that appear in the visualization, such as in the labels and legend. Technical terms often cause confusions to older adults and abbreviations can be interpreted differently due to their different historical, cultural, and educational backgrounds. This is consistent with prior work studying the usability of website and instructions for older adults~\citep{fan2018guidelines,chadwick2002web}. 

It is worth noting that DG2 raises a tension between keeping the texts concise, which often requires using technical terms and abbreviations, and using plain and easy-to-understand language, which might result in lengthy explanations. Future work should explore better ways to balance between these two aspects when designing visualizations.

\textbf{DG3: Match visual designs with older adults' expectations by understanding and following their existing knowledge and experiences}. Older adults, as any users, bring their personal experiences and knowledge when interacting with and comprehending visualizations. Thus, visualizations following their expectations can reduce their cognitive load and learning effort. 

One possible approach to implementing DG3 is to follow similar designs of typical software and tools that older adults commonly use. For example, if a map is used to visualize geolocation-related data (e.g., COVID cases in different cities), it would be desirable for the visualization to follow common visual designs (e.g., +/- icons to zoom in/out) and interaction techniques (e.g., click and drag the cursor to pan the map) in digital maps (e.g., Google Maps).

\textbf{DG4: Support trial-and-error exploration and error recovery}. \rv{Previous research found that older adults used a trial-and-error approach to learn information technology and use mobile phones~\citep{selwyn2003older,tang2013motivates,mitzner2008older,leung2012older}. Our study extends prior work and shows that trial-and-error approach was also used by older adults to recovery from errors when using \textit{interactive visualization}.} 
However, our work also shows that such trial-and-error approaches often did not help older adults recover effectively from errors. One reason was that once entering a wrong path, older adults could not easily find a way back~\citep{ziefle2005older}. One possible way to support trial-and-error exploration is to \rv{identify when older adults enter a wrong state, for example by sensing their subtle verbalization patterns (e.g., slowed-down speech and more frequent use of negative or filler words) ~\citep{fan2021older}, and promptly} help older adults return to a previous step or a ``safe'' home state so that they could easily start over again. 

Another potential reason was that the interactive visualizations never provided any timely error messages. Without informative error messages, even if participants realized that they made a mistake, they did not know what caused the problem. Consequently, they were unable to recover from it with the aimless trial-and-error approaches. This suggests that informative error messages should be provided promptly to inform older adults the potential cause of the problem and ideally a potential solution. 

% Future work should explore how older adults use a trial-and-error approach with visualizations and the challenges they encounter to better support older adults comprehend visualizations.   
% might have difficulty keeping track of the paths that they already tried and thus  because it often required them to remember outcomes of trials and backtrack steps, which could be demanding on short-term memory. As a workaround, older adults tried their familiar approaches, such as pen-and-paper. 

Finally, participants suggested another error-recovery solution, which was to provide effective instructions about how to read and interact with the visualizations. \rv{While prior research provides guidelines for creating senior-friendly instructions to help them use technology products~\citep{fan2018guidelines,tang2013motivates}, our work points out that showing instructions on a visualization may make it cluttered and negatively affects its clarity.} Thus, how to balance the number of instructions and a clean visual design is still a challenge. \rv{Future work should explore ways to balance the number of instructions and visual design, such as showing instructions on-demand.}  

%For example, it would be helpful to include comprehensive and easy-to-understand instructions, and maintaining the balance between being informative and ``clean'' for visual appearance.

\textbf{DG5: Provide responsive, reliable, and intuitive interactions}. Based on the interaction challenges described in Section~\ref{sec:challenges}, we provide six possible directions to implement this DG: 1) ensure synchronicity between user inputs and feedback (and informing users when a delay is unavoidable); 2) avoid interactions that depend on precise pointing (e.g., the densely distributed bars and lines in Figures~\ref{fig:interactivevisualizations}B and E. If precision pointing is unavoidable, consider to adopt assistive pointing techniques for older adults, such as PointAssist~\citep{hourcade2010pointassist}; 3) prevent unwanted interactions from being accidentally triggered (e.g., accidentally activated tooltips); 4) support Select when Abstract/Elaborate (e.g., tooltip information) is used so that users can choose to pin the elaborated information if needed; 5) support direction manipulation with the visual elements of the visualization instead of asking older adults to operate controls (e.g., buttons, slide bars) positioned far away from the visualization; and 6) provide multiple ways to understand data by supporting Encode and Reconfigure; \rrv{7) Tangible user interfaces have been shown to be intuitive for older adults to comprehend visualizations. For example, MeViTa was a tangible Augmented Reality (AR) based system that projected relevant medical information (e.g., side effect) beside physical medicine packages placed on a table~\cite{de2017mevita}. The combination of tangible objects (i.e., medical packages) and corresponding visualizations helped older adults understand their side effects and interactions among those medicine packages. Future work could explore tangible user interfaces to provide intuitive interactions for older adults to better comprehend visualizations.}  

\rv{Last but not least, it is worth noting that our proposed guidelines would likely make interactive visualizations more usable in general and thus benefit both older and young adults at the same time. More research is needed to tease out the general guidelines for all age groups and specific ones more applicable to older adults.}

%% file: 6-limitations-futurework.tex
\subsection{Limitations and Future Work}
% We have studied older adults' thought processes and challenges when they interacted with and comprehended interactive visualizations and presented design considerations to make interactive visualizations accessible for older adults.
% However, our findings might not cover all types of thought processes and issues that older adults may have with interactive visualizations.
% Future work should further validate and extend the findings by conducting more studies with more older adults to uncover their thought processes and challenges encountered with a broader set of interactive visualizations.

\textbf{Effects of Visualizations}. We studied how older adults interacted with COVID-19 interactive visualizations as these visualizations were critically relevant to their daily lives. %Based on participants' think-aloud data and interviews, we derived design guidelines for making interactive visualizations more accessible to older adults. 
COVID-19 data represents a typical type of temporal data that is frequently updated (e.g., on a daily basis) and contains multiple data properties (e.g., new cases, cumulative cases, death cases). Thus, our findings could potentially inform the design of the visualizations of future pandemic data for older adults. However, it remains an open question of how older adults would comprehend visualizations of other data sources and how the design guidelines might need to be adapted. 

\rv{We chose five visualizations to cover common types of interaction techniques highlighted in the literature~\citep{dimara2019interaction,yi2007toward} in our study. These interaction techniques, however, might not be equally important from older adults' perspectives. For example, some interaction techniques might be more frequently used than others. Future work should investigate the usage frequency of different interaction techniques and focus on optimizing more frequently used interaction techniques if simultaneously optimizing different ones is too costly or impossible.}

%It is, however, worth noting that the type of information conveyed in the visualizations and its relationship with older adults might affect their comprehension processes. Thus, it is worth investigating the visualizations of other topics, beyond COVID-19, to validate and enrich the design considerations.
%Based on their think-aloud verbalizations and interview feedback, we uncovered common thought processes and the challenges associated. We further derived design guidelines to make interactive visualizations more accessible to older adults. 

% explored one theme of interactive visualizations---COVID---that was relevant to older adults' current daily lives. Future work should investigate other themes of interactive visualizations that are relevant to older adults to better understand whether and to what extent the subject of visualizations might affect their comprehension process and interaction patterns.  

\textbf{Effects of Tasks}. We explored older adults' thought processes and challenges when using interactive visualizations with \textit{goal-oriented tasks}. In other words, the participants had clear goals in mind when interacting with the visualizations.
While this is a common usage scenario of visualizations, users might also explore visualizations without a specific goal in mind and perform a free-style exploration. We suspect that there might be differences in users' interaction patterns and comprehension thought processes. 
For example, there might be more open-ended trial-and-errors. Future work should explore how older adults comprehend and interact with visualizations without a specific goal to compare with and enrich the findings of this study.

\rv{We designed the tasks in our study so that they utilized key information in the COVID visualizations. Alternatively, another approach to constructing tasks is through a needs-finding study. For example, by observing how older adults actually use COVID visualizations, we might be able to derive more representative tasks for them.}

\textbf{Effects of Older Adults' Background Information}. Older adults are not a homogeneous group of users. Our study participants did not report any physical, perceptual, or cognitive impairments. As older adults' physical, perceptual, and cognitive abilities may vary widely and can affect the way how they interact with and comprehend visualizations, it is imperative to understand how older adults' varied abilities may affect how they interact with visualizations and to adapt the design guidelines accordingly. Moreover, our participants lived in North American cultures. As cultures can affect the way how people read, reason, and communicate, more research is warranted to investigate how older adults of different cultures may comprehend visualizations and how the guidelines might need to be tuned accordingly.

\textbf{Effects of Device Configurations}. Our participants joined the study remotely using their own computers in their homes. 
This setup allowed them to interact with and comprehend\st{ed} interactive visualizations on their familiar devices in their familiar environments. As a result, their thought processes and interaction patterns were more likely to reflect their natural behaviors than those observed in a controlled lab study.
That said, with this natural setup, we had limited control over their devices (e.g., screen resolution and size), internet connectivity (e.g., speed), and environmental factors (e.g., lighting condition). 
Previous research suggests that device constraints could result in problems because interactive visualizations depend\st{ing} on screen real estate and mouse-based
interaction~\citep{chittaro2006visualizing,ghose2013mobile,ghosh2003design,roberts2014visualization}.
Thus, further work should conduct more controlled lab studies to understand how device configurations (e.g., screen size, resolution, internet speed) may affect older adults' comprehension of and interaction with interactive visualizations. 

% Based on the results of two pilot study sessions, we adapted our study design so that it could be done around one hour to avoid potential fatigue for participants~\citep{TimeBudg93:online}. 
% %Specifically, we iteratively revised the interactive visualizations and tasks used for the study based on feedback from pilot studies. 
% Within this time limit, we had to make a trade-off to include the most frequently appeared interactive visualization types in our study: bar, map, and line charts.
% We acknowledge that other types of visualizations are also commonly used visualizations~\citep{borkin2013makes}.
% Future work should explore older adults' thought processes and challenges when using other types of visualizations so as to make those visualizations more accessible to older adults as well. 

\textbf{Effects of Age}. Last but not least, a natural follow-up question is whether our design guidelines are older adults specific or they would also make interactive visualizations accessible to other age groups, such as young adults. Future work could conduct similar studies with young adults and compare the similarities and differences to understand the applicability of the guidelines to other age groups. 
%With the findings from different populations, we would be able to understand the common and distinctive patterns in their thinking processes and better customize interactive visualizations to fit their thought processes. 

% The problems found by no means are complete set of problems. There might be other ones. Future work could increase the types of visualizations to uncover other potential thought processes and problems. 

%% file: 7-conclusion.tex
\section{Conclusion}
Visualization has become an important tool to make sense of ever-increasing data in this data-driven world. Thus, ensuring its accessibility to all populations is key to digital equity. In this research, we conducted think-aloud usability testing and interviews with older adults using COVID-19 interactive visualizations that were relevant to their daily lives. By analyzing participants' think-aloud verbalizations, we identified four types of thought processes \rrv{reflecting how they interacted with and} comprehended the visualizations: \textit{encoding visual information}, \textit{relating visual features to concepts}, \textit{associating concepts with existing knowledge}, and \textit{recovering from errors}
%, and \textit{comprehending the tasks}. 
These four types of thought processes confirm and extend the three-process visualization comprehension theory that was developed with \textit{static visualizations} and \textit{young adults} by examining how \textit{older adults} comprehend \textit{interactive visualizations}. 

We further uncovered the challenges that older adults encountered with each thought process. Moreover, we also highlighted the challenges they encountered with the seven common types of interaction techniques adopted in the interactive visualizations. 

Based on our findings, we present five design guidelines to make interactive visualizations more accessible to older adults: \textit{increase the legibility of visual designs}, \textit{facilitate the understanding of the affordance of visual elements by adding textual labels and using plan language}, \textit{match visual designs with users' expectations by following their existing knowledge and experiences}, \textit{support trial-and-error exploration and error recovery}, and \textit{Provide responsive, reliable, and intuitive interactions}. 

Future work should examine whether and to what extent the content of the visualizations (e.g., non-COVID related information) might affect the findings and the design guidelines. We used goal-oriented tasks to study how older adults comprehended COVID-19 interactive visualizations. It remains an open question of how the types of tasks (e.g., non-goal oriented freestyle exploration) might affect the findings. Furthermore, as older adults are not a homogeneous group, it is worth exploring how their physical and cognitive abilities as well as their culture backgrounds might affect the findings and the design guidelines. Last but not least, future work could conduct similar studies with young adults and compare the similarities and differences to understand the applicability of the guidelines to other age groups. 
%future work should further explore visualization accessibility issues for older adults with impairments as well as other types of the ``information poor'' populations who might be prone to be alienated by fast-advancing visualization techniques.         